\documentclass[12pt,preprint]{revtex4}
\pdfoutput=1

\usepackage{graphicx}
\usepackage{amsmath}
\usepackage[pdftex]{hyperref}

\usepackage[tight,nice]{units}
\usepackage{amssymb}
\usepackage[amssymb]{SIunits}
\usepackage{sistyle}

\begin{document}

\title{Modeling Reactive Wetting when Inertial Effects are Dominant}

\author{Daniel Wheeler}
\email{daniel.wheeler@nist.gov}
\author{James A. Warren}
\author{William J.  Boettinger}

\affiliation{Metallurgy Division, Materials Science and Engineering Laboratory,
  National Institute of Standards and Technology, Gaithersburg, MD 20899, USA}

\date{\today}

\begin{abstract}

  Recent experimental studies of molten metal droplets wetting high temperature
  reactive substrates have established that the majority of triple-line motion
  occurs when inertial effects are dominant. In light of these studies, this
  paper investigates wetting and spreading on reactive substrates when inertial
  effects are dominant using a thermodynamically derived, diffuse interface model
  of a binary, three-phase material. The liquid-vapor transition is modeled using
  a van der Waals diffuse interface approach, while the solid-fluid transition is
  modeled using a phase field approach. The results from the simulations
  demonstrate an $O \left( t^{-\nicefrac{1}{2}} \right)$ spreading rate during
  the inertial regime and oscillations in the triple-line position when the metal
  droplet transitions from inertial to diffusive spreading. It is found that the
  spreading extent is reduced by enhancing dissolution by manipulating the
  initial liquid composition.  The results from the model exhibit good
  qualitative and quantitative agreement with a number of recent experimental
  studies of high-temperature droplet spreading, particularly experiments of
  copper droplets spreading on silicon substrates.  Analysis of the numerical
  data from the model suggests that the extent and rate of spreading is regulated
  by the spreading coefficient calculated from a force balance based on a
  plausible definition of the instantaneous interface energies. A number of
  contemporary publications have discussed the likely dissipation mechanism in
  spreading droplets. Thus, we examine the dissipation mechanism using the
  entropy-production field and determine that dissipation primarily occurs in the
  locality of the triple-line region during the inertial stage, but extends along
  the solid-liquid interface region during the diffusive stage.

\end{abstract}


\maketitle

\section{Introduction}

Characterizations of metal alloys wetting and spreading on dissolving substrates
typically assume that inertial effects are not dominant or that the majority of
dissipation is due to viscous forces~\cite{Warren19983247, villanueva:056313,
  ISI:000266518600024, ISI:000272111800009}.
In many respects this seems an entirely reasonable approach since the majority of
experiments do not capture the early time behavior when inertial effects are
dominant, but focus on the late-stage spreading when chemical-diffusion dominates
and substrate dissolution occurs. Typically, experimental studies measure only
slow spreading on the order of seconds or even minutes for millimeter-sized metal
droplets consistent with diffusion-dominated spreading~\cite{ISI:000073465700012,
  Warren19983247, ISI:000079337600004, ISI:000225453200025}.  However, using
improved techniques, a number of recent experiments~\cite{Grigorenko,
  ISI:000225453200025, ISI:000258573900040, ISI:000271268400034} capture the
rapid early-stage spreading and demonstrate that the spreading duration is
consistent with the inertial time scale~\cite{ISI:000250621900080}. The
variations in experimental findings can be attributed to differences in substrate
temperature, composition of the vapor phase influencing substrate oxidation,
contact mechanisms between the substrate and molten droplet, camera shutter
speed, as well as other factors~\cite{ISI:000225453200025}. An often important
aspect of managing these factors is arresting the formation of a substrate ridge
on which the triple line becomes attached, which can retard spreading
considerably~\cite{ISI:000073465700012}.

The spreading droplet is often characterized in terms of a velocity versus
contact angle relationship where the velocity is scaled using the instantaneous
Capillary number, $ \operatorname{Ca}^* = U^* \nu / \gamma$, where $U^*$ is the
instantaneous spreading speed, $\nu$ is the liquid viscosity and $\gamma$ is the
liquid-vapor interface energy. Saiz~\textit{et al.}  postulated that the
dissipation mechanism may not be due to viscous forces as previously
understood~\cite{ISI:000165747000010, ISI:000225453200025}.  Clearly, in cases
where the dissipation mechanism is not due to viscous effects,
$\operatorname{Ca}$ is no longer a useful quantity for characterizing the
spreading and an alternative parameter is required. An effective ``triple-line
friction'' derived from molecular kinetics theory is suggested by Saiz~\textit{et
  al.} that is independent of viscosity but still dependent on interface energy
and the contact angle. A number of recent experimental
studies~\cite{ISI:000250621900080} clearly show that a large proportion of the
spreading is characterized entirely by the inertial time scale ($t_i = \sqrt{\rho
  R_0^3 /\gamma}$, where $\rho$ is the liquid density and $R_0$ is the drop
radius) with $U \sim t^{-\nicefrac{1}{2}}$, which is much faster than typical
viscous spreading laws~\cite{PhysRevE.69.016301}. Furthermore, molecular dynamics
studies of Ag-Ni and Ag-Cu systems seem to confirm the $t^{-\nicefrac{1}{2}}$
dependence of the spreading rate even for relatively small
droplets~\cite{WebbIII20053163, 0953-8984-21-46-464135}.

This paper employs a diffuse interface method in order to analyze the issues
surrounding the inertial spreading regime and dissipation mechanism discussed
above. The diffuse interface approach implicitly includes a wide range of
phenomena and as such does not require a posited relationship between spreading
rate and contact angle~\cite{CambridgeJournals:15613}.  Villanueva~\textit{et
  al.}~\cite{ISI:000272111800009} used a diffuse interface method to model
reactive wetting and clearly identified two separate spreading regimes: an
initial viscous regime and a subsequent diffusive
regime~\cite{villanueva:056313}. The viscous regime demonstrated excellent
agreement with standard viscous spreading laws. Further work by these
authors~\cite{ISI:000272111800009} employed the same model to examine the effects
of dissolution on spreading by first recovering the non-dissolutive hydrodynamic
limit as a base state. In the viscous regime they found the spreading to be
independent of the diffusion coefficient, but accelerated in the diffusive regime
as the diffusion coefficient is increased. This paper outlines a similar process
using the initial liquid concentration to vary the driving force for dissolution,
while maintaining a constant diffusion coefficient. The general consensus of the
literature is that inertial spreading occurs more slowly in systems that exhibit
dissolution than in immiscible systems that do not exhibit
dissolution~\cite{Yin20063561, Warren19983247}. However, this is contradicted by
a number of experiments for saturated and pure liquids that show that the
spreading can be on a similar time scale under certain experimental
conditions~\cite{ISI:000258573900040, ISI:000225453200025}.

The work of Villanueva~\textit{et al.}~\cite{ISI:000272111800009} considers
droplets that do not exhibit inertial effects due to the small drop size, which
is limited by the requirement of having a narrow interface ($\approx
1$~\nano\metre). In contrast to reference~\cite{ISI:000272111800009}, this work
sacrifices the realistic interface width in an attempt to model a system that
exhibits inertial effects. Due to the drop size restrictions, the inertial time
scale used in Villanueva~\textit{et al.}  is $t_i \approx
\unit{\num{6e-11}}{\second}$ and the capillary time scale, $t_c = \nu R_0 /\gamma
\approx \unit{\num{2e-11}}{\second}$. At these values, the extent of spreading
during the inertial stage is limited and the characteristic inertial effects are
suppressed by viscous forces.  The Ohnesorge number, given by $\operatorname{Oh}
= t_c / t_i$, quantifies the relative importance of inertial and viscous
effects. Typically, millimeter-sized metal droplets are highly inertial in nature
with $\operatorname{Oh} \approx\num{1e-3}$. Characteristic inertial effects, such
as triple-line position oscillations and large droplet curvature variations, are
reduced for $\operatorname{Oh} > 0.01$ and eliminated for $\operatorname{Oh} >
1$~\cite{ISI:A1997YE01200009}. In Villanueva~\textit{et al.}, $\operatorname{Oh}
\approx 0.3$ and in this work $\operatorname{Oh} \approx \num{6e-3}$.


Jacqmin makes an extensive study of the role of the diffuse interface method,
specifically for a Cahn-Hilliard--van der Waals system (CHW), in relieving the
stress singularity that occurs for classical sharp interface
methods~\cite{CambridgeJournals:15613}.  Since the interface is diffuse, the CHW
does not require an explicit alteration to the no-slip boundary condition to
allow for triple-line slip. Jacqmin demonstrates that the CHW has the same far
field and macroscopic behavior as classical hydrodynamic models of slip.  Thus,
in diffuse interface models that include hydrodynamics there is no need to define
a slip length.  The interface width determines both an effective slip length and
the concentration profiles within the diffuse interface associated with
adjustments to adsorption and desorption; these factors affect the evolution of
the system in subtle ways. There is no exact expression relating interface width
and the effective slip length, however, $\lambda = \delta / 2 R_0$ is suggested
as a good rule of thumb in Ding and Spelt~\cite{CambridgeJournals:964080}, where
$\lambda$ is the dimensionless effective slip length for a diffuse interface
model. It is claimed that the slip length can be as large as
\unit{50}{\nano\metre}~\cite{ISI:000077188100038}, which is close to the chosen
interface width in the present work, although the drop radius is only
\unit{1}{\micro\metre}. The slip length is found by Ding and Spelt to influence
the onset of oscillations that occur when the droplet transitions from the
inertial stage to the diffusive stage. The critical value of
$\operatorname{Re}^*$ for which oscillations occur is reduced with decreasing
$\lambda$. Hocking and Davis~\cite{ISI:000178373700001} have demonstrated that
there is no simple relationship between contact angle and velocity when the
approach to equilibrium becomes oscillatory, which seems to be the case in a
number of experimental and numerical studies of millimeter-sized
droplets~\cite{Grigorenko, ISI:000258573900040, CambridgeJournals:964080,
  ISI:A1997YE01200009}.

The code used for the numerical analysis in this paper is developed using the
FiPy PDE solver~\cite{guyer:6}. Details of how to install FiPy as well as the
reactive wetting code used here are given on the FiPy web
site~\cite{reactiveWettingInstallation}. The numerical analysis and figures
presented in this paper can be reproduced with the open source tools
available. The underlying linear solvers and parallel capabilities are provided
by the Trilinos tool suite~\cite{1089021}.

In the following section the governing equations are presented followed by a
discussion of the associated dimensionless parameters in
section~\ref{section:dimensionless}. Results from the numerical solution of the
governing equations outlined are presented in
section~\ref{section:results}. Section~\ref{section:discussion} analyzes the
results in the context of previous work and ends with a discussion of the
dissipation mechanism. Section~\ref{section:conclusion} presents the
conclusions. Appendix~\ref{appendix:derivation} derives the governing equations
presented in section~\ref{section:equations}, while
appendix~\ref{appendix:numericalApproach} presents details of the numerical
methods.

\section{Governing Equations \label{section:equations}}

In this section, the final forms of the governing equations are presented along
with the associated thermodynamic parameters and functions. The full derivation
of the governing equations is described in
appendix~\ref{appendix:derivation}. The system consists of a three phase (solid,
liquid and vapor) binary alloy. The liquid-vapor system is modeled as a two
component van der Waals fluid, while the solid-fluid system is modeled with a
phase field description. The density field acts as the order parameter for the
liquid-vapor transition. Thus, the system is fully characterized by the
spatio-temporal evolution of the mass density of component 1, $\rho_1$, the mass
density of component 2, $\rho_2$, the phase field, $\phi$, as well as the
barycentric velocity field $\vec{u}$, as determined through the momentum
equation. The three dimensional equations are reduced to two dimensions by
imposing cylindrical symmetry about $r=0$. The initial configuration consists of
a spherical droplet with a radius of $1$~$\mu$m tangent to a solid substrate
surrounded by a vapor.  The incompressible approximation is not made in this work
for numerical reasons outlined in appendix~\ref{appendix:numericalApproach}; all
the phases are compressible. The solid is modeled as a very viscous fluid as in
previous phase field reactive wetting studies~\cite{villanueva:056313,
  ISI:000272111800009}. As the total mass density, $\rho=\rho_1+\rho_2$, appears
so frequently in the equations, it is more convenient to use $\rho$ and $\rho_2$
as the independent density variables. For economy in notation, we write spatial
derivatives $\partial_i\equiv
\partial/\partial x_i$, $\partial_i^2 \equiv \partial^2/\partial x_i^2$ and
require that repeated indices are summed, unless otherwise indicated. Note that
although the equations are solved with cylindrical symmetry, the equations are
presented in the following Cartesian forms:
\subsubsection{Continuity}
\begin{equation}
  \label{eqn:continuity}
  \frac{\partial \rho}{\partial t}+\partial_j\left(\rho u_j \right) =0.
\end{equation}
\subsubsection{Diffusion}
\begin{equation}
  \label{eqn:diffusion}
  \frac{\partial \rho_2}{\partial t}+\partial_j\left(\rho_2 u_j\right)=
  \partial_j\left( \frac{M}{T}\partial_j\left(\mu_2^{NC}-\mu_1^{NC}\right)\right).
\end{equation}
\subsubsection{Phase}
\begin{equation}
  \label{eqn:phase}
  \frac{\partial \phi}{\partial t} + u_j\partial_j\phi
= \epsilon_{\phi} M_{\phi} \partial_j^2 \phi - \frac{M_{\phi}}{T}
  \frac{\partial f}{ \partial \phi}
\end{equation}
\subsubsection{Momentum}
\begin{figure}
  {\hfill\scalebox{0.75}{\includegraphics{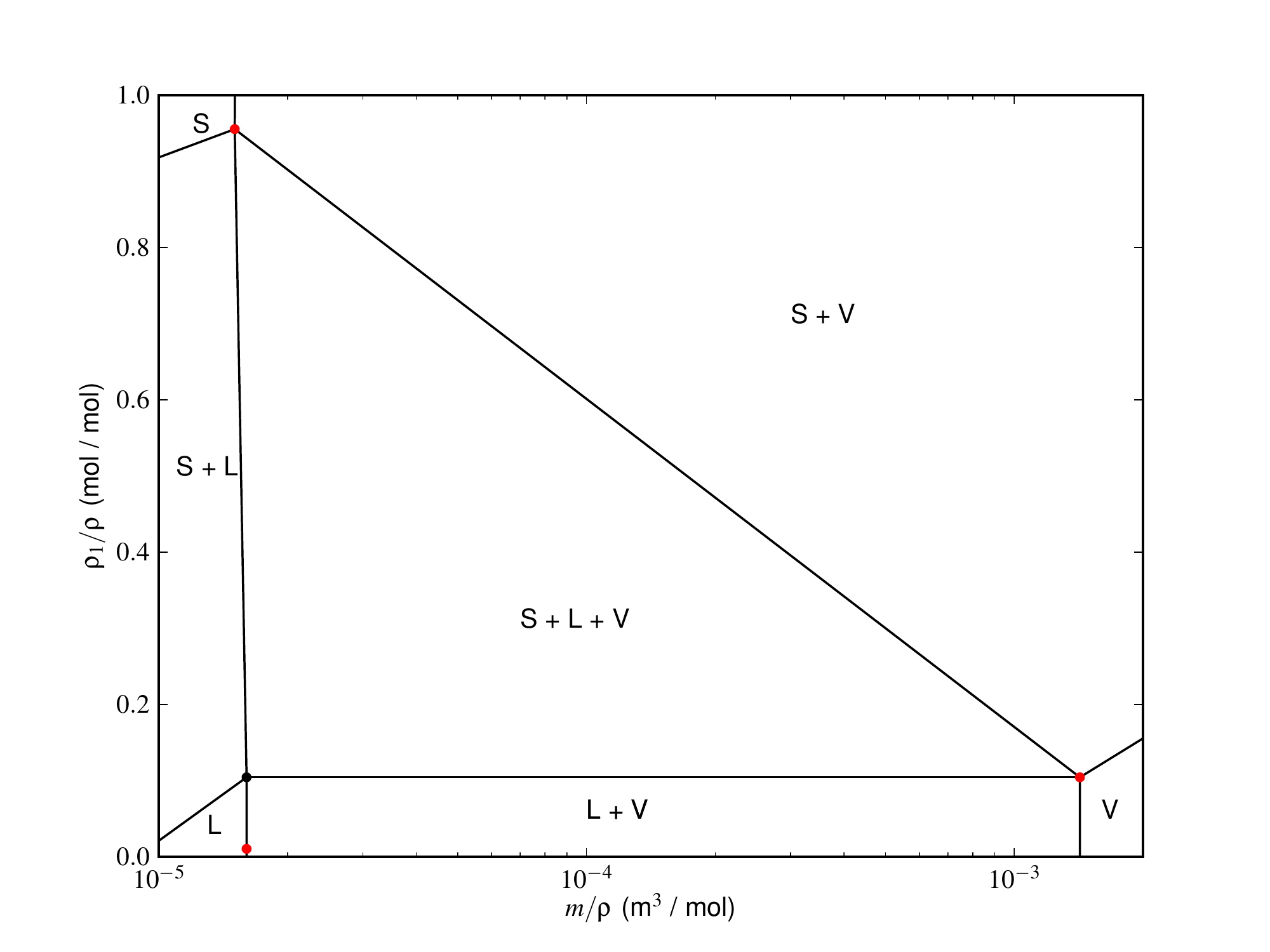}}\hfill}
  \caption{The phase diagram for the system of parameters presented in
    Table~\ref{tab:2Dsimulation}. Each region represents a possible equilibrium
    state for a mixture of solid (S), liquid (L) and vapor (V) phases. The red
    dots represent the initial conditions for the $\xi=0.1$ simulation discussed
    in section~\ref{section:results}. The black dot marks the liquid equilibrium
    condition. The liquid and vapor phases are thick in component 1 while the
    solid phase is thick in component 2.}
\label{fig:phaseDiagram}
\end{figure}

\begin{equation}
\begin{split}
  \label{eqn:momentum2}
  \frac{\partial \left( \rho u_i\right) }{\partial t} +
  \partial_j\left( \rho u_iu_j \right) &=
  \partial_j\left( \nu \left[ \partial_j u_i + \partial_i u_j \right] \right)
  \\ &- \rho_1 \partial_i \mu_1^{NC} - \rho_2 \partial_i \mu_2^{NC} + 
  \left( \epsilon_{\phi} T \partial_j^2 \phi -
  \frac{\partial f}{ \partial \phi} \right) \partial_i \phi
\end{split}
\end{equation}
where $u_i$ is a velocity component, $T$ is the temperature and $M=\bar{M} \rho_1
\rho_2 / \rho^2$ is the chemical mobility, which is proportional to the
diffusivity, $D$, as outlined in Eq.~\eqref{eqn:diffusionCoeff}. The values of
$\bar{M}$ and $\nu$ vary from the solid to the fluid phases with the
interpolation scheme chosen to be
\begin{equation}
  \bar{M} = \bar{M}_s^{\psi} \bar{M}_f^{1- \psi}
  \label{eqn:mbar}
\end{equation}
and
\begin{equation}
  \nu = \nu_s^{\psi} \nu_f^{1- \psi}
  \label{eqn:nu}
\end{equation}
where $\psi = \phi^a$ with $a=4$. The values used in the simulations for
$\bar{M}_s$, $\bar{M}_f$, $\nu_s$ and $\nu_f$ are in
Table~\ref{tab:2Dsimulation}. The choice of $a$ is discussed in
subsection~\ref{subsection:interface}. The free energy per unit volume is
postulated to have the form~\cite{plischke1994equilibrium},
\[
f = p\left( \phi \right) f_s + \left(1 - p \left( \phi \right) \right)
f_f + W \phi^2 \left( 1 - \phi \right)^2
\]
where $W$ is the phase field barrier height and
$p(\phi)=\phi^3(10-15\phi+6\phi^2)$ represents a smoothed step function common in
phase field models~\cite{BoettingerReview:2002}. The free energies per unit
volume in the separate fluid and solid phases are given by,
\begin{equation}
  f_f = \frac{e_1 \rho_1^2}{m^2} + \frac{e_{12} \rho_1 \rho_2}{m^2}
  + \frac{e_2 \rho_2^2}{m^2} + \frac{ R T }{ m } \left[ \rho_1
    \ln{\rho_1} + \rho_2 \ln{\rho_2} - \rho \ln{\left( m - \bar{v}
        \rho \right)} \right]
\label{eqn:fluidfreeenergy}
\end{equation}
and
\begin{equation}
  f_s = \frac{A_1 \rho_1}{m} +
  \frac{A_2 \rho_2}{m} +
  \frac{R T}{m} \left( \rho_1 \ln \rho_1 + \rho_2 \ln \rho_2 - \rho \ln \rho \right)
  + \frac{B}{\rho m} \left( \rho_s^{\text{ref}} - \rho \right)^2
\label{eqn:solidFreeEnergy}
\end{equation}
where $m$ is the molecular weight (assumed to be equal for each component), R is
the gas constant, $\bar{v}$ is the exclusion volume due to the finite size of the
atoms, $B$ is the solid compressibility, $\rho_s^{\text{ref}}$ is a reference
density for the solid and the $e_i \rho_i / m$ are the free energy contributions
per unit mole due to intermolecular attraction in the van der Waals model. The
$A_1$ and $A_2$ are temperature dependent parameters related to the heat of
fusion between the solid and fluid phases. Along with the free energy, the
specification of the pressure and the non-classical chemical potentials are
required to fully define the system,
\begin{eqnarray}
  P &=& \rho_1 \frac{\partial f}{\partial \rho_1} + \rho_2
  \frac{\partial f}{\partial \rho_2} - f \label{eqn:P} \\
  \mu_1^{NC} &=& \frac{\partial f}{\partial \rho_1} - \epsilon_1 T
  \partial_j^2 \rho_1 \label{eqn:mu1} \\
  \mu_2^{NC} &=& \frac{\partial f}{\partial \rho_2} - \epsilon_2 T
  \partial_j^2 \rho_2 \label{eqn:mu2}
\end{eqnarray}
where $\epsilon_1$ and $\epsilon_2$ are free energy gradient coefficients. The
parameter values for Eqs.~\eqref{eqn:continuity} to~\eqref{eqn:mu2} are presented
in Table~\ref{tab:2Dsimulation}. The corresponding isothermal phase diagram for
the molar fraction of component 1 verses the molar volume is displayed in
figure~\ref{fig:phaseDiagram}.

Eqs.~\eqref{eqn:continuity}--\eqref{eqn:momentum2} are solved using a
cell-centered, collocated finite-volume (FV) scheme. The solution algorithm uses
a fully coupled Krylov solver with Picard non-linear updates using $\rho_1$,
$\rho_2$, $\phi$ and $\vec{u}$ as the independent variables. Further discussion
of the numerical approach is given in appendix~\ref{appendix:numericalApproach}.

\section{Dimensionless Equations and Timescales\label{section:dimensionless}}

It is useful for the purposes of analysis and completeness to clearly present the
various dimensionless numbers and time scales that arise from solving
Eqs.~\eqref{eqn:continuity}~\eqref{eqn:diffusion}~\eqref{eqn:phase} and
~\eqref{eqn:momentum2} in the context of spreading droplets. The dimensionless
forms of Eqs.~\eqref{eqn:diffusion} and~\eqref{eqn:momentum2} are given by,
\begin{equation}
  \label{eqn:diffusion-dimensionless}
  \frac{\partial \rho_2 }{\partial t} +
  \partial_j \left( u_j \rho_2 \right)
  =
  \frac{1}{\operatorname{Pe}} \partial_j \left( \frac{\rho_1 \rho_2}{\rho^2} \partial_j
    \left( \mu_2 - \mu_1 -
      \operatorname{Q}
    \partial_k^2 \left( \rho_2 - \rho_1 \right)
    \right) \right)
\end{equation}
and
\begin{equation}
  \frac{\partial \left( \rho u_i \right) }{\partial t} +
  \partial_j \left( \rho u_i u_j \right) =
  \frac{1}{\operatorname{Re}} \partial_j \left( \partial_j u_i + \partial_i u_j
  \right) - 
  \frac{1}{\operatorname{Ma}^2} \partial_i P + \frac{1}{\operatorname{We}} \left( \rho_1 \partial_i \partial_j^2 \rho_1 + \rho_2 \partial_i \partial_j^2 \rho_2
    - \tilde{\epsilon}_{\phi} \partial_i \phi \partial_j^2  \phi \right)
  \label{eqn:momentum-dimensionless}
\end{equation}
where the variables and operators are now dimensionless (the analysis of
Eqs.~\eqref{eqn:continuity} and~\eqref{eqn:phase} is not particularly revealing
and is omitted). For completeness, all the time scales referred to in this paper
are displayed in table~\ref{tab:timescales} as a prerequisite for presenting the
dimensionless numbers in table~\ref{tab:dimensionlessnumbers}.
It should be noted that in table~\ref{tab:timescales}, $U^*=U^* \left( t \right)$
is the instantaneous spreading speed and $U$ is a fixed spreading speed posited
\textit{a priori}.

The time scale $t_{\text{diff}}$ represents the time required for the
solid-liquid interface to move a distance $\delta$ due to diffusion mediated
melting or freezing.  The expression for $t_{\text{diff}}=\delta^2 / 4 K^2 D_f$
is determined using an error function based similarity solution
(see~\cite{kirkendallBeam}) where $K$ is the solution to
\[
K + \left( \frac{X_1^l - X_1^{l, \text{equ}} }{ X_1^s - X_1^l } \right)
\frac{\exp{ \left( -K^2 \right) }} {1 - \text{erf} \left( K \right)}
\frac{1}{\sqrt{\pi}} = 0
\]
and the chemical diffusion coefficient in the fluid, $D_f$, is defined by
\begin{equation}
  D_f = \frac{\bar{M}_f R}{m \rho_l^{\text{equ}}} = \unit{\num{9.58e-10}}{\square\meter\per\second}
  \label{eqn:diffusionCoeff}
\end{equation}
If we substitute $R_0$ for $\delta$ in the expression for $t_{\text{diff}}$, a
rough estimate is obtained for complete equilibration of the system. Since
$t_{\text{diff}} \gg t_i$, the motion of the solid-liquid interface is negligible
for a simulation that is both computationally feasible and adequately resolves
the inertial time scale. The motion of the solid interface due to dissolution is
controlled by both diffusion ($t_{\text{diff}}$) and boundary kinetics
(represented by $t_\phi$). Here $t_\phi\ll t_i$, thus dissolution will be limited
by diffusion rather than boundary kinetics. Additionally, solid interface motion
due to hydrodynamic effects is negligible because the solid viscosity is chosen
such that $t_s \gg t_i$ where $t_s$ represents the time scale for discernible
motion of the solid.

Table~\ref{tab:dimensionlessnumbers} presents the dimensionless numbers in terms
of their constituent time scales where appropriate.  Note that there are now two
separate expressions for both the Reynolds number and the Capillary number based
on $U$ and $U^*$. By making an informed choice for the value of $U$, estimates
are obtained for the likely values of the dimensionless numbers when using
$U^*$. Here, $U=R_0 / t_i = \unit{\num{5.08e1}}{\meter \per \second}$ is selected
based on the spreading rate for a system that is dominated by inertial
effects. The values of $\operatorname{Oh}$, $\operatorname{Re}$ and
$\operatorname{Pe}$ in table~\ref{tab:dimensionlessnumbers} all indicate that the
interface energy and inertial forces dominate over viscous and diffusive
forces. Since $\operatorname{We}=1$, the interface energy and inertial forces are
of approximately equivalent magnitude. Small values of $\operatorname{Oh}$ are
representative of many experimental systems of technical interest: for example,
$\operatorname{Oh} \approx \num{2e-3}$ for a millimeter sized droplet of copper
and $\operatorname{Oh} \approx \num{2e-2}$ for a micrometer-sized drop of lead.

Other dimensionless numbers (included for completeness) in
table~\ref{tab:dimensionlessnumbers} include the Mach number,
$\operatorname{Ma}$, which requires a definition for the speed of sound in the
liquid, given by~\cite{landauLifshitz},
\begin{equation*}
  c = \left.\sqrt{\frac{\partial P}{\partial \rho}} \right|_{ \rho_l^{\text{equ}}} = \unit{\num{8.89e2}}{\meter\per\second}
\end{equation*}
and $\operatorname{Q}$, which represents the ratio between interface and internal forces in the
liquid droplet, but has not been identified in the literature by the authors.

\section{Results \label{section:results}}

In this section, we explore the rate and extent of droplet spreading based on
variations in the initial liquid concentration and the Ohnesorge number. The
initial liquid concentration determines the driving force for dissolution, while
manipulating the Ohnesorge number influences the impact of inertial effects on
spreading. The results presented here will provide the basis for comparison with
other authors' work in section~\ref{section:discussion}.

The extent of dissolution is established by decreasing the initial value of the
liquid concentration, $X_1^{l}$, requiring the solid to dissolve in order to
restore $X_1^{l}$ to its equilibrium value, $X_1^{l, \text{equ}}$. Explicitly, we
set
\begin{equation}
\begin{split}
  \label{eq:initialConcentration}
  X_1^l \left(t=0\right) &= \left(1 - \xi \right) X_1^{l,\text{equ}} \\
  \rho_l \left(t=0\right) &= \rho_l^{\text{equ}}
\end{split}
\end{equation}
where $\xi$ defines a measure of the magnitude of the driving force for
dissolution ($\xi < 0$ induces freezing). When $\xi = 0$, the system has no
potential for dissolution, similar to pure hydrodynamic spreading where surface
tension forcing dominates and interface motion is due only to convection as phase
change is negligible. In this limit, comparisons can be made with simpler
spreading models and power laws. In addition to the hydrodynamic case ($\xi=0$),
simulations were conducted with values of $\xi=0.5$ and $\xi=0.9$.

\begin{figure}
  {\hfill\scalebox{0.75}{\includegraphics{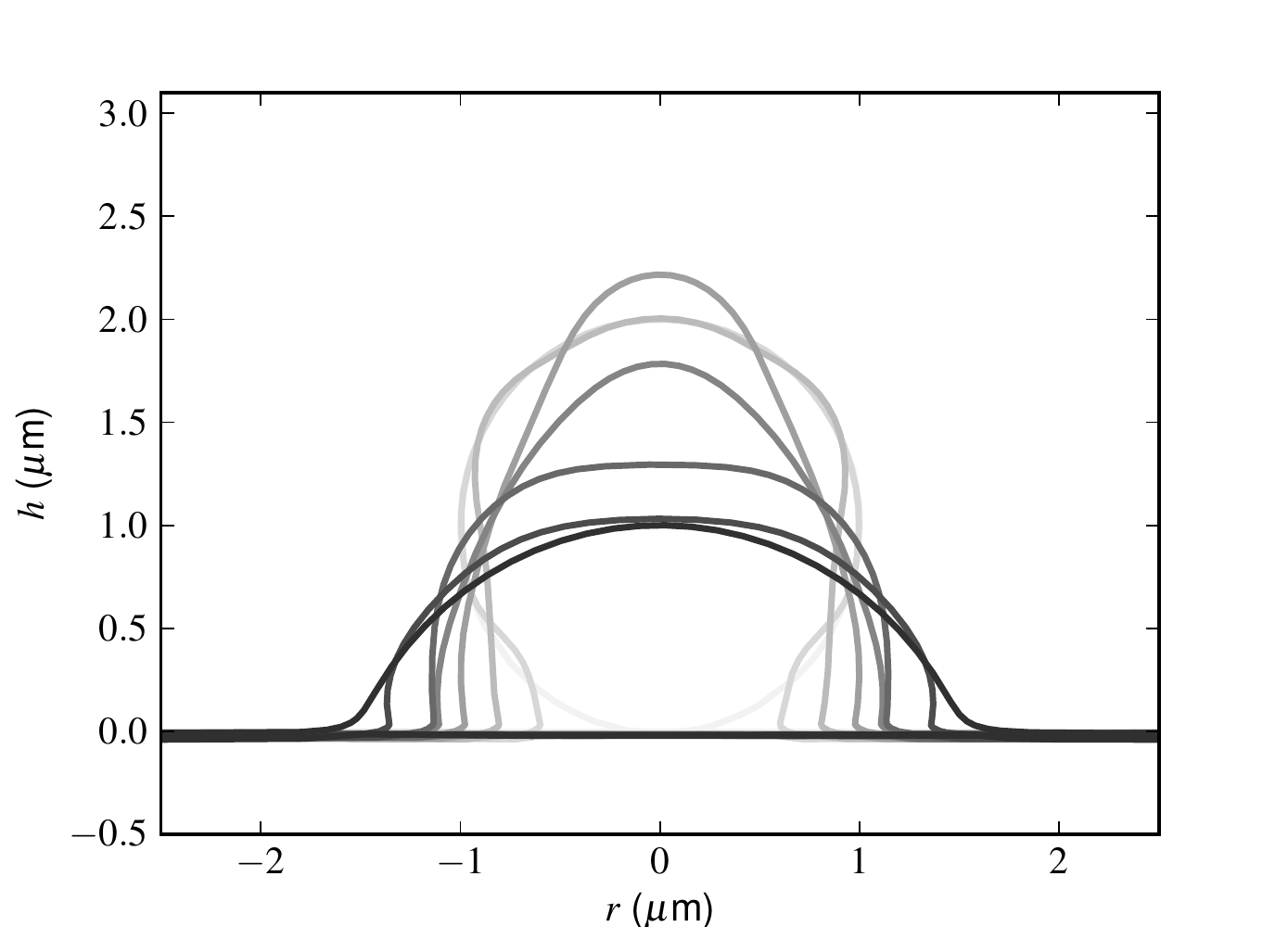}}\hfill}
  \caption{Sequential configurations of the liquid-vapor and solid-fluid
    interfaces for $\xi=0$ and $\operatorname{Oh}=5.7 \times 10^{-3}$ with darker
    tones indicating later times. The curves demonstrate the extreme inertial
    effects on the droplet. The droplet starts as a sphere in tangent contact
    with the substrate. The drop height then rises considerably as the capillary
    wave initiated from the triple line arrives at the top of the
    droplet. Although large amplitude ($\approx R_0/5$) oscillations occur in the
    triple-line position, the largest contact angle oscillation is only
    $\approx\unit{0.03\pi}{\radian}$.}
\label{fig:density1}
\end{figure}

Figure~\ref{fig:density1} demonstrates the highly inertial nature of the
spreading dynamics. Upon initiation of the simulation, pressure waves appear at
the interface regions and travel through the interior of the droplet, but then
disperse quickly. Simultaneously, triple-line motion begins with a rapid change
in the local contact angle, but without any discernible motion elsewhere on the
drop interface. This initiates the most noticeable feature of the spreading: a
capillary wave propagates from the triple line along the liquid-vapor interface,
initiating the onset of the triple-line motion and progressing to the top of the
droplet, causing a rapid rise in the drop height. The wave then travels back to
the triple-line location while the droplet completes the majority of the
spreading, with both events having a duration that corresponds to $\approx 2
t_i$. During this interval, the triple-line motion is monotonic and without
interruption. On return to the triple-line location, the wave induces a reversal
in the triple-line motion. Subsequent waves induce further reversals in the
triple-line motion and the drop height with a period of $\approx 2 t_i$. The
amplitude of the oscillations diminishes in the manner of an under-damped
oscillator, completing approximately 5 or 6 full cycles before ceasing
entirely. Subsequently, very slow monotonic spreading occurs with the
liquid-vapor interface appearing to have almost constant curvature.

\begin{figure}
  {\hfill\scalebox{0.75}{\includegraphics{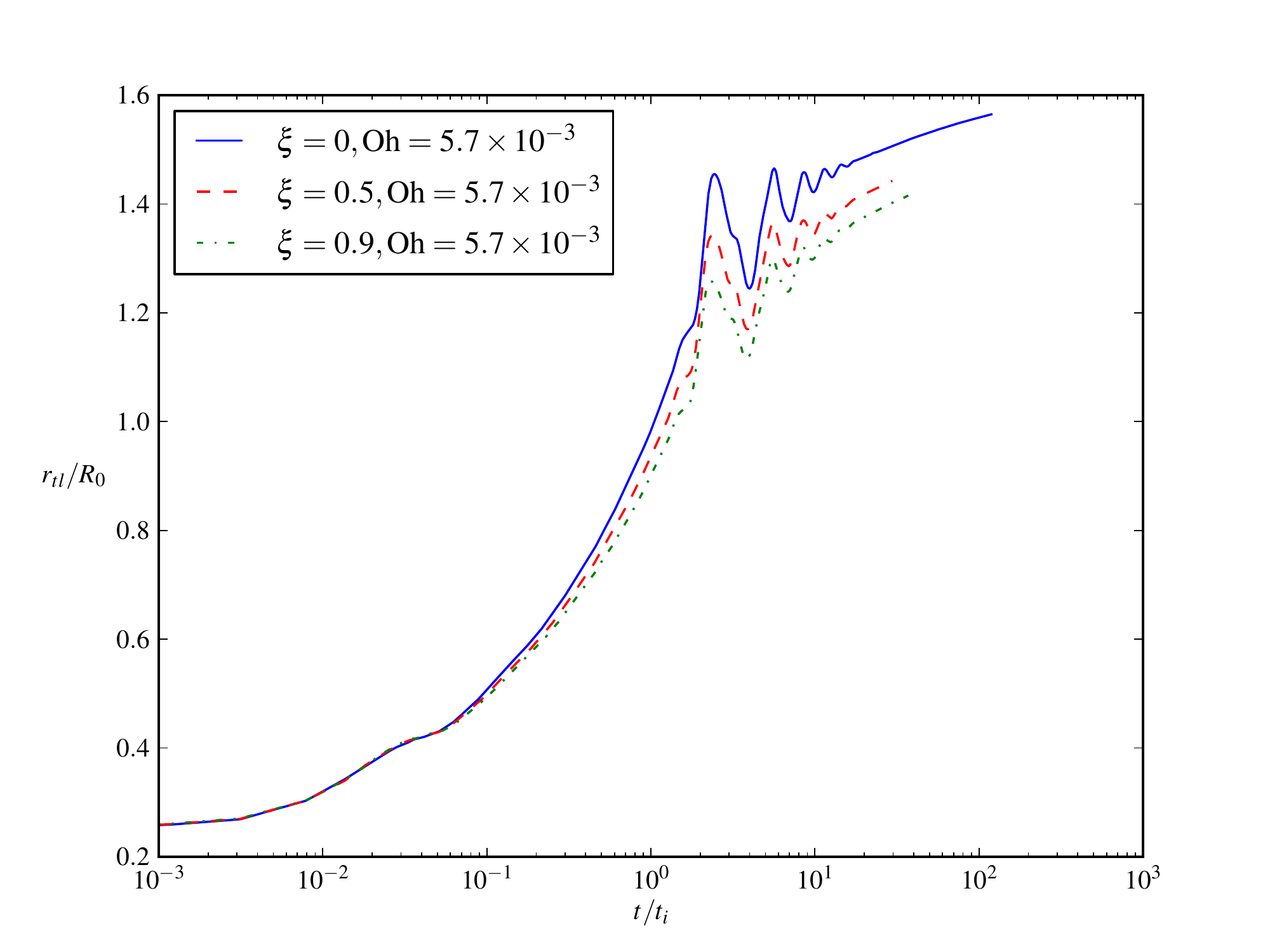}}\hfill}
  \caption{The spreading radius versus time for various values of $\xi$ with
    $\operatorname{Oh}=5.7 \times 10^{-3}$. As $\xi$ increases, the spreading
    rate and extent of spreading is slightly reduced.}
  \label{fig:dropRadius}
\end{figure}

\begin{figure}
  {\hfill\scalebox{0.75}{\includegraphics{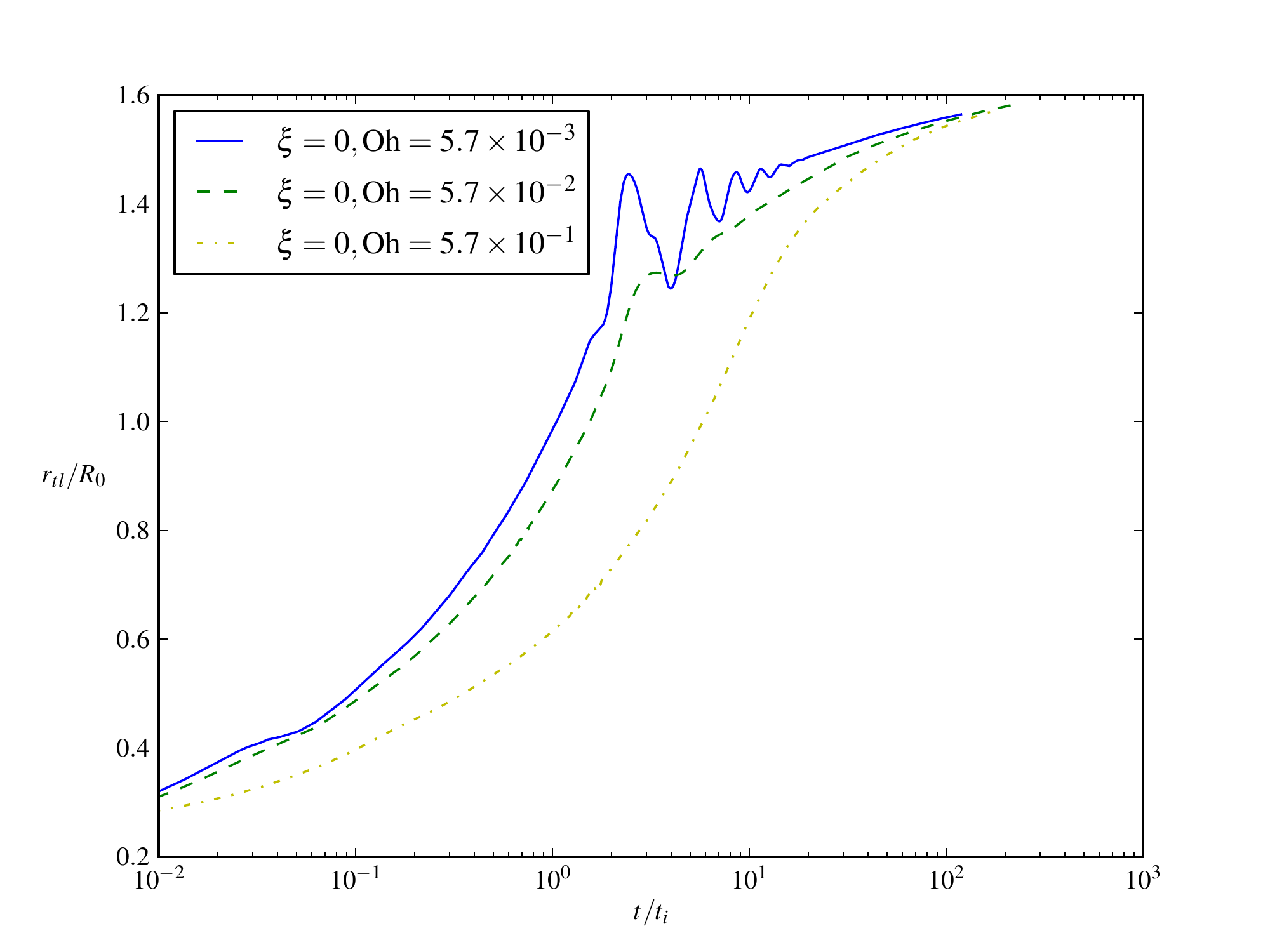}}\hfill}
  \caption{The spreading radius, $r_{tl}$, versus time for various values of
    $\operatorname{Oh}$ with $\xi=0$. The oscillations are eliminated for the
    largest value of $\operatorname{Oh}$.}
  \label{fig:dropRadius-nu}
\end{figure}

Figures~\ref{fig:dropRadius} and~\ref{fig:dropRadius-nu} display the scaled
radial position of the triple-line, $r_{tl} / R_0$, against the scaled time for
varying values of $\xi$ and $\operatorname{Oh}$. The two intervals of fast and
slow monotonic spreading can clearly be seen as well as the intervening period of
oscillatory spreading as discussed in the previous paragraph. Increasing $\xi$
reduces the extent of spreading slightly, while increasing $\operatorname{Oh}$
eliminates the oscillations entirely and considerably reduces the spreading
rate. In each of these cases, the actual amount of substrate dissolution is
negligible (the solid-fluid interface moves less than $\delta / 5$) due to the
large disparity between the dissolutive and inertial time scales as discussed in
section~\ref{section:dimensionless}. In figure~\ref{fig:dropRadius}, at early
times ($t < 0.1 t_i$), the value of $\xi$ has no impact on the spreading, but at
later times ($t > 0.1 t_i$) the curves diverge. When $t> 10 t_i$, the curves stop
diverging and seem to remain at a fixed distance apart.  Increasing $\xi$ not
only results in a slight reduction in the extent of spreading, but also results
in a notable reduction in the amplitude of the oscillations. These factors
indicate that there is a seemingly modest decrease in the driving force for
spreading with increasing $\xi$. In figure~\ref{fig:dropRadius-nu}, the
$\operatorname{Oh}=5.7\times10^{-1}$ curve diverges from the other curves at very
early times and has a greatly diminished spreading rate. Eventually, the curves
become coincident at late times when the spreading is free of observable inertial
manifestations for all values of $\operatorname{Oh}$.

\begin{figure}
  {\hfill\scalebox{0.75}{\includegraphics{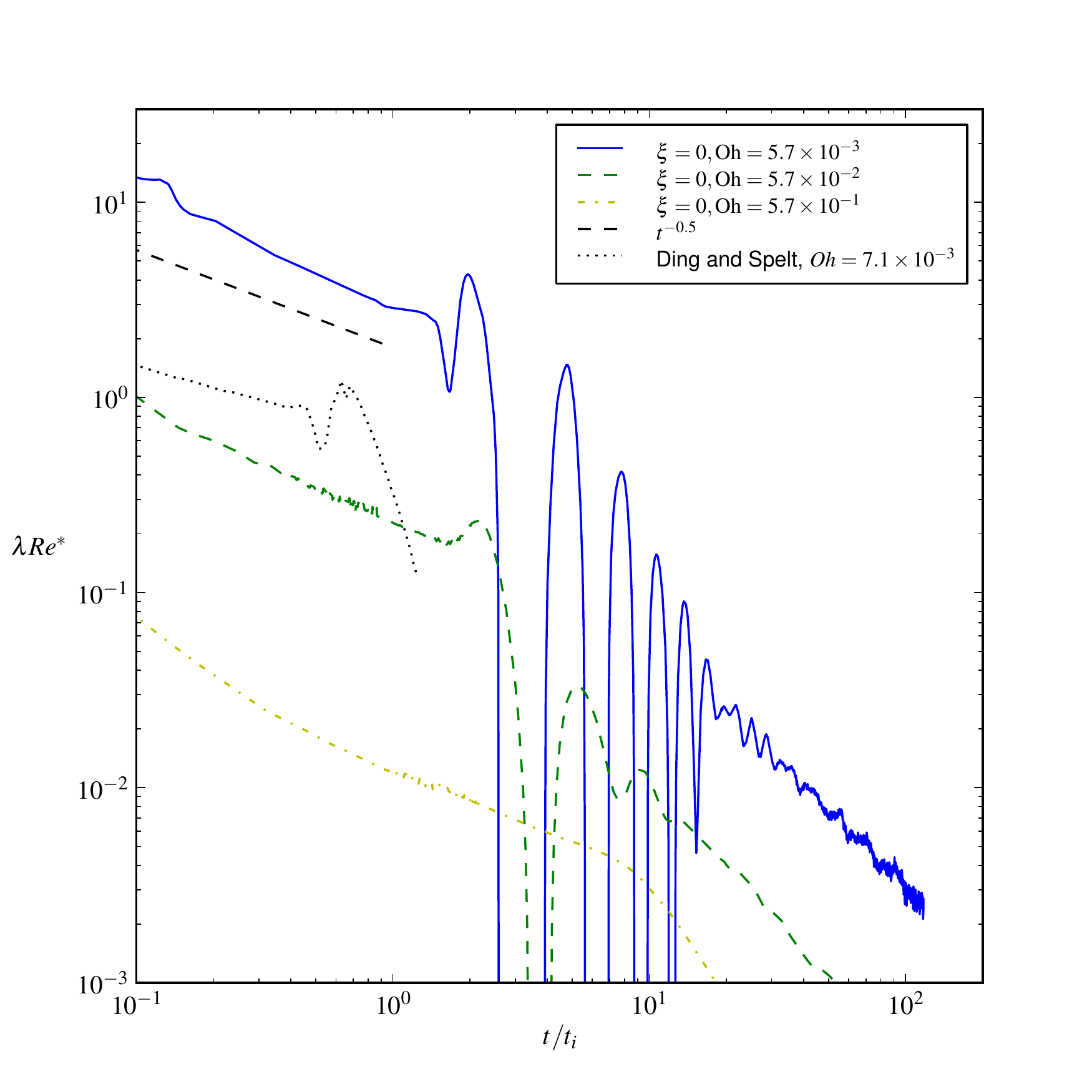}}\hfill}
  \caption{The dimensionless spreading rate against the dimensionless time with
    varying $\operatorname{Oh}$ and $\xi=0$. The spreading occurs in three
    distinct intervals.. The sign changes in the blue curve correspond to the
    triple-line oscillations during the transition from the inertial to the
    diffusive regime.}
\label{fig:triplePointVelocityVtime}
\end{figure}

In order to compare with other models, the radial position results presented in
figure~\ref{fig:dropRadius-nu} are presented using a scaled spreading velocity in
figure~\ref{fig:triplePointVelocityVtime}.  The spreading velocity is scaled
using a Reynolds number, $\lambda \operatorname{Re}^*$ ($\lambda = \delta / 2
R_0$), based on the interface width, $\delta$, rather than using a standard
Reynolds number based on the initial drop radius,
$R_0$~\cite{CambridgeJournals:964080}. The spreading velocity data used in
figure~\ref{fig:triplePointVelocityVtime} is smoothed to remove noise on the
order of a grid spacing, the details of which are described in
appendix~\ref{appendix:numericalApproach}. The sign changes in the blue curve,
when $t_i<t<10t_i$, correspond to the triple-line oscillations seen in
figure~\ref{fig:dropRadius-nu}. The oscillations lie between intervals with
monotonically decreasing spreading velocity. The $\operatorname{Oh}=\num{5.7e-1}$
(yellow) curve exhibits a fairly steady decrease in velocity and then a much
sharper reduction when $t \approx 10t_i$, which corresponds to a slope change in
frigure~\ref{fig:dropRadius-nu}. Note that the $\operatorname{Oh}$ values for
simulations presented in figure~\ref{fig:triplePointVelocityVtime} are
manipulated by changing the value of $\nu_f$ only, and thus, a corresponding
figure with no scaling for the spreading velocity would show only slight
differences between the vertical positions of the curves.

\begin{figure}
  {\hfill\scalebox{0.75}{\includegraphics{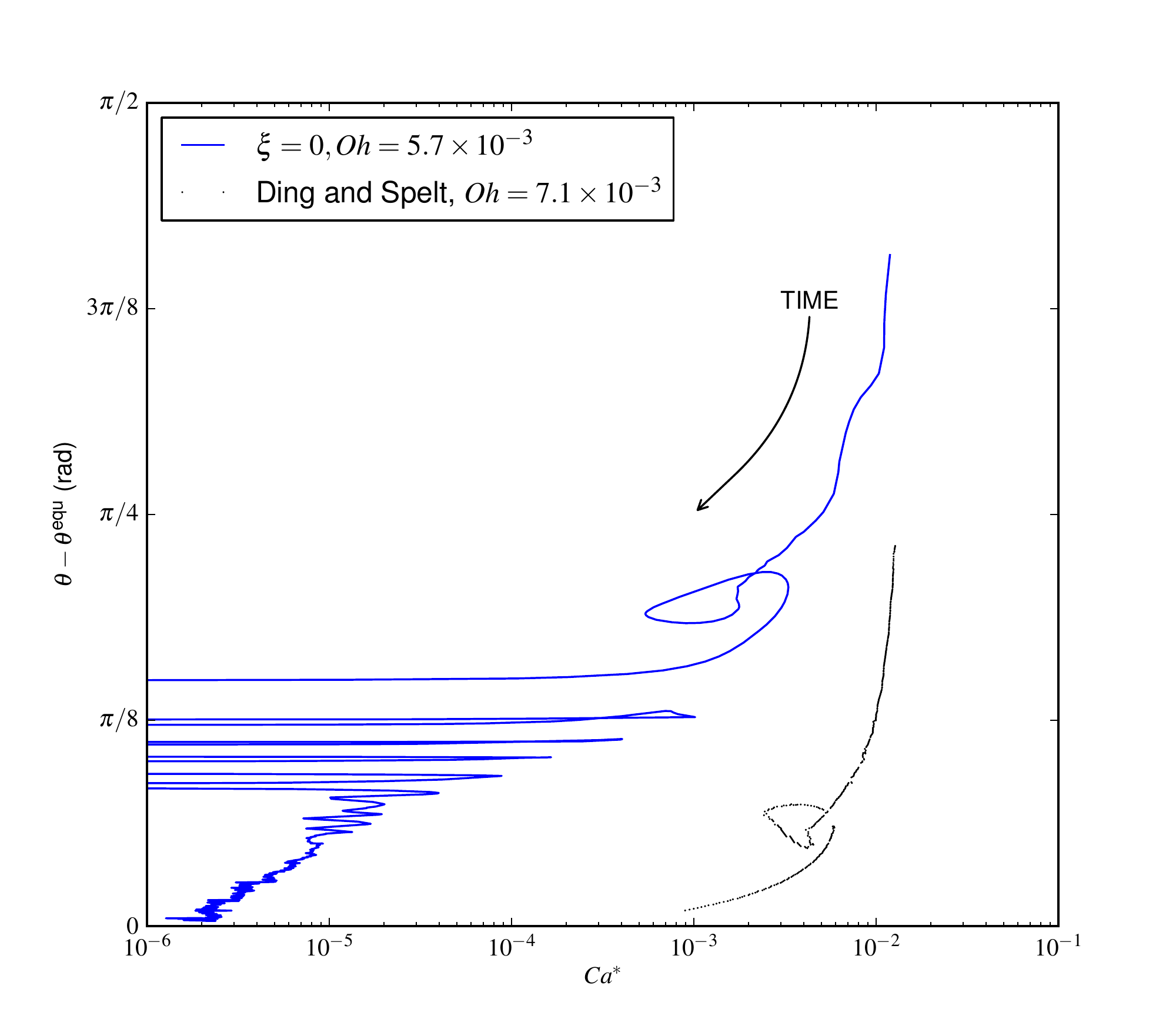}}\hfill}
  \caption{The observed contact angle against $\operatorname{Ca}^*$ for $\xi=0$
    and $\operatorname{Oh}=\num{5.7e-3}$.}
\label{fig:CaVAngle}
\end{figure}

Figure~\ref{fig:CaVAngle} displays the apparent contact angle, $\theta$, against
the Capillary number for $\operatorname{Oh}=\num{5.7e-3}$ and $\xi=0$
demonstrating the convergence of $\theta$ to the nominal flat-interface,
equilibrium contact angle, $\theta^{\text{equ}}$. The angle $\theta$ is
calculated using techniques similar to those described in Villanueva~\textit{et
  al.}~\cite{ISI:000272111800009}. Although $\theta$ exhibits a hysteresis loop,
it remains relatively steady during the period of oscillatory spreading and only
varies by $\approx \unit{0.03\pi}{\radian}$ for the largest oscillation.

\section{Discussion\label{section:discussion}}

\subsection{Comparison with other models}

At early times, the flow is dominated by inertia and comparisons with theories of
spreading on flat, non-reactive substrates are fruitful.  Indeed, an analytical
spreading rate for the inertial regime can be derived, see Biance~\textit{et
  al.}~\cite{PhysRevE.69.016301}, and is given by $t^{-\nicefrac{1}{2}}$. In
figure~\ref{fig:triplePointVelocityVtime}, the slope of this power law (black
dashed line) shows reasonable agreement with the $\operatorname{Oh}=\num{5.7e-3}$
(blue curve) during the inertial regime. The vertical position of the black
dashed line is selected to enable easy comparison with the blue curve.

In the work of Ding and Spelt~\cite{CambridgeJournals:964080}, phase field and
level set models of a spreading droplet are compared for a range of Ohnesorge
numbers ($\num{7.1e-3} \le \operatorname{Oh} \le \num{2.8e-1}$) making it a
useful study for comparing with our work. The black dotted curve in
figure~\ref{fig:triplePointVelocityVtime} is a digitized curve of the lowest
value of $\operatorname{Oh}$ simulated in Ding and Spelt. This particular
simulation is selected for display here as it manifests the most pronounced
oscillations. They simulate droplets with an initial contact angle of
$\unit{\pi/3}{\radian}$ and an equilibrium contact angle of $\unit{\pi /
  18}{\radian}$ using an effective dimensionless slip length of $\lambda = 0.01$
($\lambda=0.05$ in our work).  Despite these differences, the overall motion of
the droplets agrees well qualitatively for droplets with similar Ohnesorge
numbers, although triple line motion was not seen to reverse direction in their
work. In figure~\ref{fig:CaVAngle}, the contact angle experiences a hysteresis
loop in a similar fashion to the work of Ding and Spelt, which is reproduced in
the black dotted curve.

It has been conjectured~\cite{CambridgeJournals:964080, ISI:000178373700001} that
the value of $\lambda \operatorname{Re}^*$ controls whether or not the spreading
becomes oscillatory. In the simulations presented here, $\lambda
\operatorname{Re}^*$ varies between 1 and 10 for the lowest value of
$\operatorname{Oh}$, but this is harder to determine for experimental
systems. Hydrodynamic analysis of experimental data results in a slip length that
can vary substantially for different materials (typically between
\unit{1}{\nano\metre} and
\unit{100}{\nano\metre}~\cite{ISI:000225453200025}). Using these bounds, a
typical millimeter sized metal drop results in $ 0.01 < \lambda
\operatorname{Re}^* < 1$ assuming a spreading rate of \unit{1}{\metre\per\second}
(in this work the spreading rate is $\approx\unit{50}{\metre\per\second}$). It is
interesting to note that for values of $\lambda \operatorname{Re}^* < 0.1$, no
oscillatory motion was seen in the work by Ding and
Spelt~\cite{CambridgeJournals:964080}. In Schiaffino and
Sorin~\cite{ISI:A1997YE01200009} it is experimentally determined that the
transition between under-damped oscillations to over-damped decay (no
oscillations) occurs as $\operatorname{Oh}$ increases above
$1\times10^{-2}$. This is seemingly confirmed in figure~\ref{fig:dropRadius-nu}
where the curve that corresponds to $\operatorname{Oh}=5.7 \times 10^{-3}$ has
multiple oscillations, while the curve for $\operatorname{Oh}=5.7 \times 10^{-1}$
has no oscillations.

\subsection{Comparison with experiments}

\begin{figure}
  {\hfill\scalebox{0.75}{\includegraphics{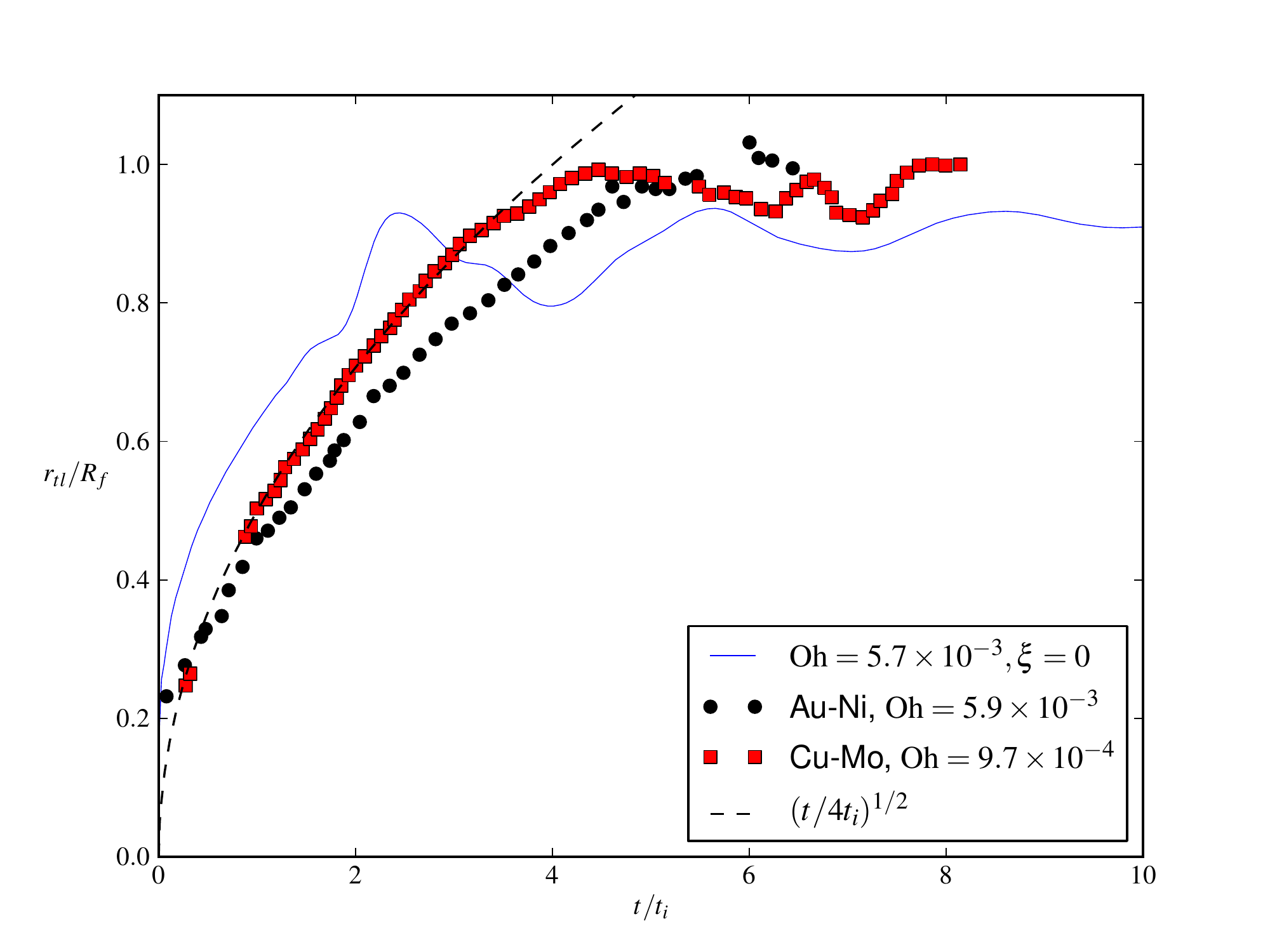}}\hfill}
  \caption{The radial position of the triple line scaled against the final radial
    position, $R_f$, against time (scaled with $t_i$) for $\xi=0$, Au-Ni
    experimental results and Cu-Mo experimental results. The experimental results
    are digitized from Saiz~\textit{et al.}~\cite{ISI:000225453200025,
      ISI:000250621900080}. The inertial time scale, $t_i$ for the Au-Ni system
    is calculated using $\rho =
    \unit{1.1\times10^{4}}{\kilogram\per\cubic\metre}$, $\gamma =
    \unit{1.0}{\joule\per\square\metre}$, $R_0 = \unit{1 \times
      10^{-3}}{\metre}$. The inertial time scale for the Cu-Mo system is
    calculated using $\rho = \unit{8.9\times10^{3}}{\kilogram\per\cubic\metre}$,
    $\gamma = \unit{1.3}{\joule\per\square\metre}$ and $R_0 = \unit{1 \times
      10^{-3}}{\metre}$. The value of $t_i$ is $\unit{1.9\times10^{-8}}{\second}$
    for this work, $\unit{3.4\times10^{-3}}{\second}$ for the Au-Ni system and
    $\unit{2.6\times10^{-3}}{\second}$ for the Cu-Mo system. This figure shows
    the reasonable agreement between the simulation and experimental data when
    scaled by the inertial time scale and the agreement with the $(t/ 4
    t_i)^{\nicefrac{1}{2}}$ spreading rate.}
\label{fig:nature}
\end{figure}

\begin{figure}
  {\hfill\scalebox{0.75}{\includegraphics{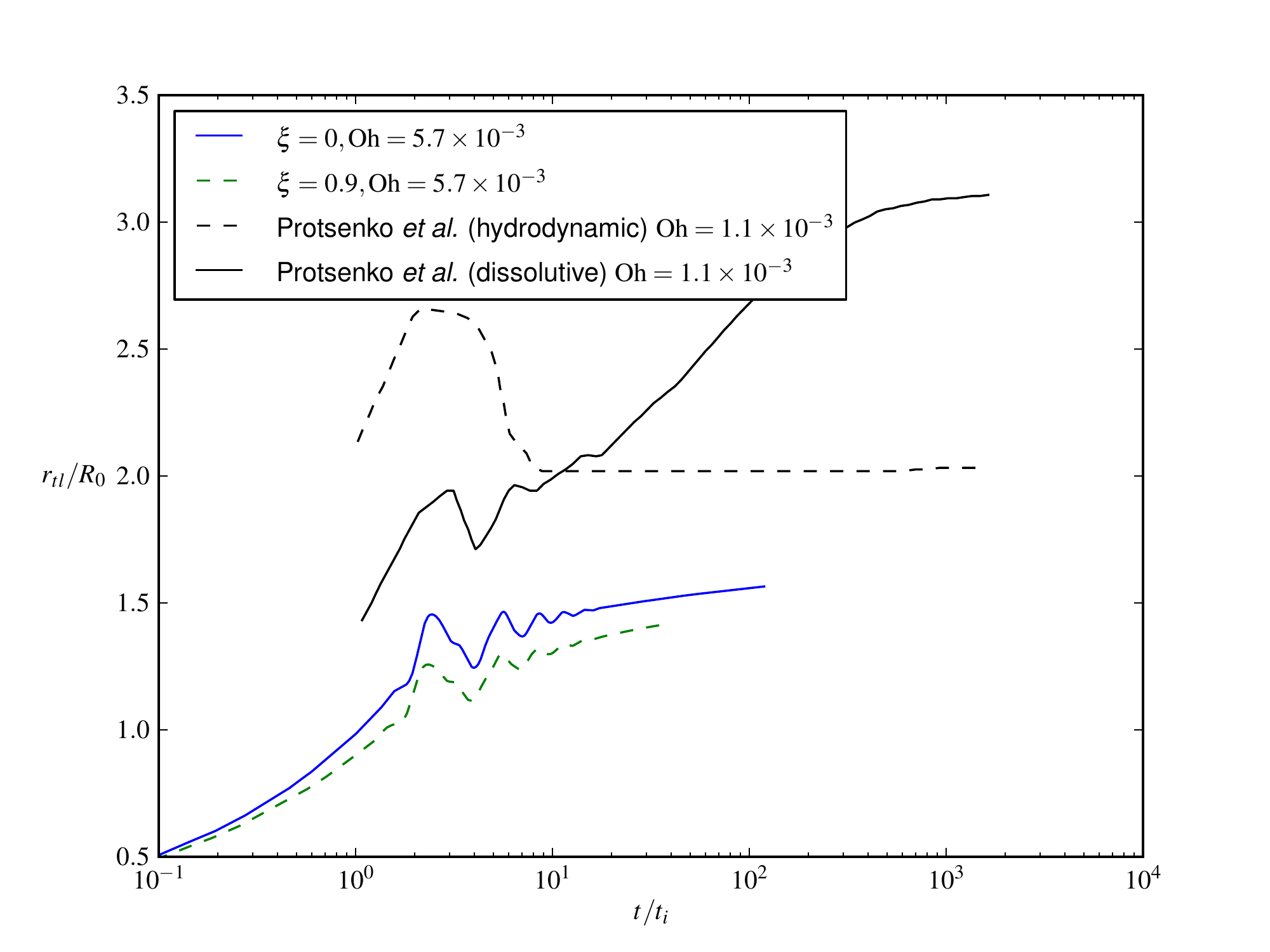}}\hfill}
  \caption{The spreading radius versus time for $\xi=0$ and $\xi=0.9$. The black
    curves are Cu-Si experiments digitized from Protosenko~\textit{et al.}. The
    inertial time scale, $t_i$, for the Cu-Si system is calculated using $\rho =
    \unit{8.9\times10^{3}}{\kilogram\per\cubic\metre}$, $\gamma =
    \unit{1.3}{\joule\per\square\metre}$ and $R_0 = \unit{8.2 \times
      10^{-4}}{\metre}$. The value of $t_i$ is $\unit{1.9\times10^{-8}}{\second}$
    for this work and $\unit{2.0\times10^{-3}}{\second}$ for the Cu-Si system }
\label{fig:protsenko}
\end{figure}

In figures~\ref{fig:nature} and~\ref{fig:protsenko} the triple-line radial
position results from the present work for $\operatorname{Oh}=\num{5.7e-3}$ are
compared with experimental results from Saiz and
Tomsia~\cite{ISI:000225453200025, ISI:000250621900080} and Protsenko~\textit{et
  al.}~\cite{ISI:000258573900040}.  These experiments are conducted at a high
temperature (\unit{1100}{\celsius}) and exhibit fast spreading, which is either
absent or undocumented in many other reactive wetting
experiments~\cite{ISI:000073465700009}. In Saiz and Tomsia, the experimental
results are for Au and Cu droplets with an initial radius of
\unit{1}{\milli\metre} spreading on Ni and Mo substrates, respectively, while in
Protsenko~\textit{et al.} the experiments are for Cu droplets of a similar size
spreading on Si substrates.  The reasonable quantitative agreement between the
experimental and simulation results in figures~\ref{fig:nature}
and~\ref{fig:protsenko} (within $\approx20\%$ for the Cu-Mo combination) suggests
that the spreading in the experimental systems is predominantly inertial in
nature~\cite{ISI:000250621900080}.

The Cu on Mo spreading in figure~\ref{fig:nature} indicates oscillatory behavior
at the end of the inertial regime, although there are only a handful of data
points supporting this claim. Also, since the period of any oscillations is
likely to be $\approx 2t_i$, a much greater duration of experimental data is
required for confirmation. The dissolutive case (black solid curve) in
figure~\ref{fig:protsenko} clearly demonstrates oscillations of a similar period,
amplitude and duration to the simulation results presented here as well as a
contact angle hysteresis (not shown). It should be noted that oscillatory
spreading also occurs in other systems such as water droplets on
glass~\cite{ISI:A1997YE01200009}.

\subsection{Non-equilibrium interface energy
  analysis \label{subsection:interface}}

The driving force for spreading on a planar substrate is often characterized by the
spreading coefficient given by,
\begin{equation}
  S^{\text{equ}} \left( t \right) = \gamma_{sv}^{\text{equ}}  - \left( \gamma_{sl}^{\text{equ}}  +
    \gamma_{lv}^{\text{equ}}   \cos{\theta} \left( t
    \right) \right)
\label{eqn:spreadingCoefficientEqu}
\end{equation}
where the $\gamma^{\text{equ}}$ are equilibrium values of the interface energies
and $\theta$ is the observed contact angle. The utility of
Eq.~\eqref{eqn:spreadingCoefficientEqu} is clearly limited to circumstances where
the interface energies remain close to their equilibrium values during
spreading. A number of authors~\cite{ISI:A1974T150700009, ISI:000073465700009,
  ISI:000255456800015, ISI:000271268400034} have suggested that this limitation
may be overcome by replacing the equilibrium interface energies with their
instantaneous values in Eq.~\eqref{eqn:spreadingCoefficientEqu}.  This yields a
new spreading coefficient
\begin{equation}
  \tilde{S} \left( t \right) = \tilde{\gamma}_{sv} \left( t \right) - \left( \tilde{\gamma}_{sl} \left( t \right) +
    \tilde{\gamma}_{lv} \left( t \right)  \cos{\theta} \left( t
    \right) \right)
\label{eqn:spreadingCoefficient}
\end{equation}
where the $\tilde{\gamma}$ are instantaneous interface energies.  In principal,
the use of $\tilde{\gamma}$ rather than $\gamma^{\text{equ}}$ provides a more
accurate description of the driving force for spreading, particularly in the case
where the timescale for spreading, $t_i$, is much faster than the interface
equilibration timescale, $t_{\text{diff}}$.  Since the solid-fluid interface
remains planar over the time scales of interest in the simulations, using a
horizontal force balance alone and ignoring the vertical imbalance when deriving
Eq.~\eqref{eqn:spreadingCoefficient} can be viewed as a reasonable assumption. An
alternative expression to Eq.~\eqref{eqn:spreadingCoefficient} can be derived if
the solid-fluid interface is non-planar using a more general Neumann's triangle
horizontal and vertical force balance.  In the following discussion, the
expression used to calculate the $\tilde{\gamma}$ is described and then $S$ is
used to analyze the influence of $\xi$ on the spreading dynamics.

\begin{figure}
  {\hfill\scalebox{0.75}{\includegraphics{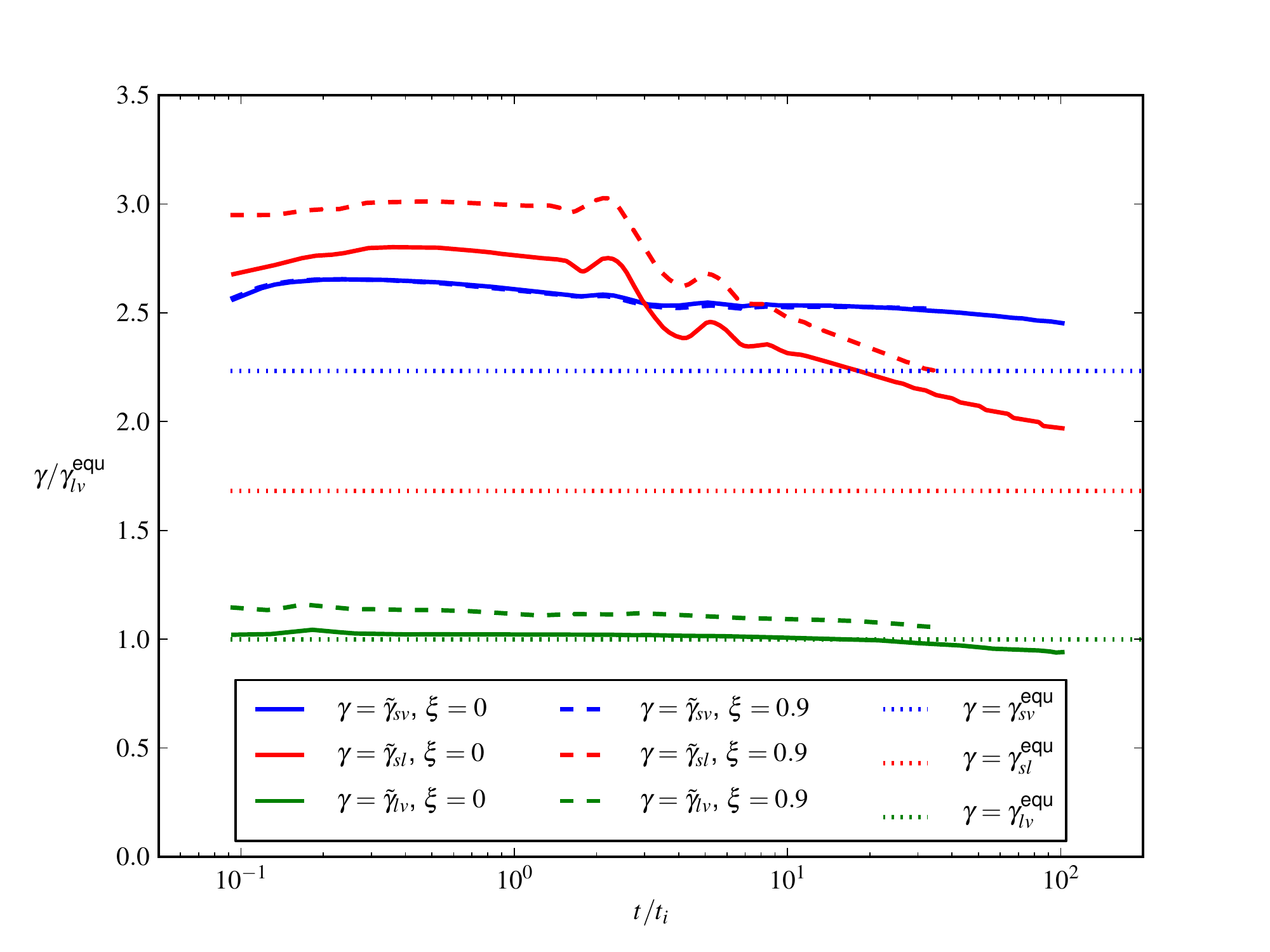}}\hfill}
  \caption{The instantaneous interface energies $\tilde{\gamma}$ plotted against
    time for $\xi=0$ and $\xi=0.9$. Both $\tilde{\gamma}_{lv}$ and
    $\tilde{\gamma}_{sl}$ are larger for the $\xi=0.9$ curve.}
  \label{fig:gamma-205}
\end{figure}

It is a substantial advantage of our approach that we are able to develop an
explicit expression for the instantaneous interface energies, allowing us to test
the utility of $\tilde S$ as a metric for spreading. In order to calculate
$\tilde S$ using the results of the present calculations, we begin with two
equivalent expressions for the equilibrium energy of a planar interface:
\begin{equation}
\begin{array}{ll}
  \gamma &= \int_{-\infty}^\infty \left[ \epsilon_1 T |\nabla \rho_1|^2 + \epsilon_2 T |\nabla \rho_2|^2
    + \epsilon_{\phi} T |\nabla \phi|^2 \right] dl  \\
 &=  2 \int_{-\infty}^\infty 
  \left[ 
    f - f^{\infty} - \mu_1^{\infty} \left( \rho_1 - \rho_1^{\infty} \right)  - \mu_2^{\infty} \left( \rho_2 - \rho_2^{\infty} \right) 
  \right] dl
\end{array}
\label{eqn:surfaceTensionIntegraleq}
\end{equation}
where the $\infty$ superscript represents the value in the far field, and all the
fields have equilibrium profiles. The equivalence of the expressions in
Eq.~\eqref{eqn:surfaceTensionIntegraleq} can be demonstrated by first writing
down the Euler-Lagrange equation derived from the free energy functional in
Eq.~\eqref{eqn:functional} with additional Lagrange multiplier terms for the
conservation of both species and then integrating once. We now assert that a
plausible measure of the instantaneous interface energy is
\begin{equation}
  \tilde{\gamma} \left( t \right) = \int_l \left[ \epsilon_1 T |\nabla \rho_1|^2 + \epsilon_2 T |\nabla \rho_2|^2
    + \epsilon_{\phi} T |\nabla \phi|^2 \right] dl
\label{eqn:surfaceTensionIntegral}
\end{equation}
where $l$ is a line segment that both intersects and is normal to the interface
being measured with $\int_l dl > \delta$. All fields in
Eq.~\eqref{eqn:surfaceTensionIntegral} are measured at time $t$.  In general, the
quantity $\tilde{\gamma}$ is a useful heuristic when the gradients are confined
to the interface region. The numerical integration of
Eq.~\eqref{eqn:surfaceTensionIntegral} is conducted at a distance of $2\delta$
from the triple-line location perpendicular to each local interface over a
distance of $1.5\delta$. The integration points on the respective interfaces are
chosen to be as near to the triple-line location as possible while avoiding the
large variations in the value of $\tilde{\gamma}$ that occur close to the
triple-line location~\cite{ISI:000272111800009}. Clearly, we could have defined
another instantaneous interface energy as,
\begin{equation}
\gamma^*(t) = 2 \int 
\left[ 
  f - f^{\infty} - \mu_1^{\infty} \left( \rho_1 - \rho_1^{\infty} \right) - \mu_2^{\infty} \left( \rho_2 - \rho_2^{\infty} \right) 
\right] dl
  \label{eqn:surfaceTensionIntegralAlt}
\end{equation}
As one approaches equilibrium $\gamma^*\rightarrow\tilde\gamma$, but dynamically
the quantities are different.  It would appear that $\gamma^*$ is less useful
than $\tilde\gamma$, as $\gamma^*$ requires the fields to be near the far field
(equilibrium) values at the integration limits extremes for the value to ``make
sense'' as an interface excess quantity. It is instructive to observe the
$\tilde{\gamma}$ behavior over time (see figure~\ref{fig:gamma-205}).  The values
of $\tilde\gamma$ differ substantially from their equilibrium values for most of
the simulation.  The $\tilde{\gamma}_{sv}$ appear independent of $\xi$, which is
a reasonable expectation, as $\xi$ sets the liquid concentration.  Increasing
$\xi$ results in an increase in both $\tilde{\gamma}_{lv}$
and$\tilde{\gamma}_{sl}$. In figure~\ref{fig:gamma-205} large oscillations can be
observed in the solid-liquid interface energy (red curve). These oscillations are
due to the spatially varying values of $\tilde{\gamma}_{sl}$ along the
solid-liquid interface in conjunction with the oscillations in the
$\tilde{\gamma}_{sl}$ integration line location moving in unison with the
triple-line location during the oscillatory phase of motion.

\begin{figure}
  {\hfill\scalebox{0.75}{\includegraphics{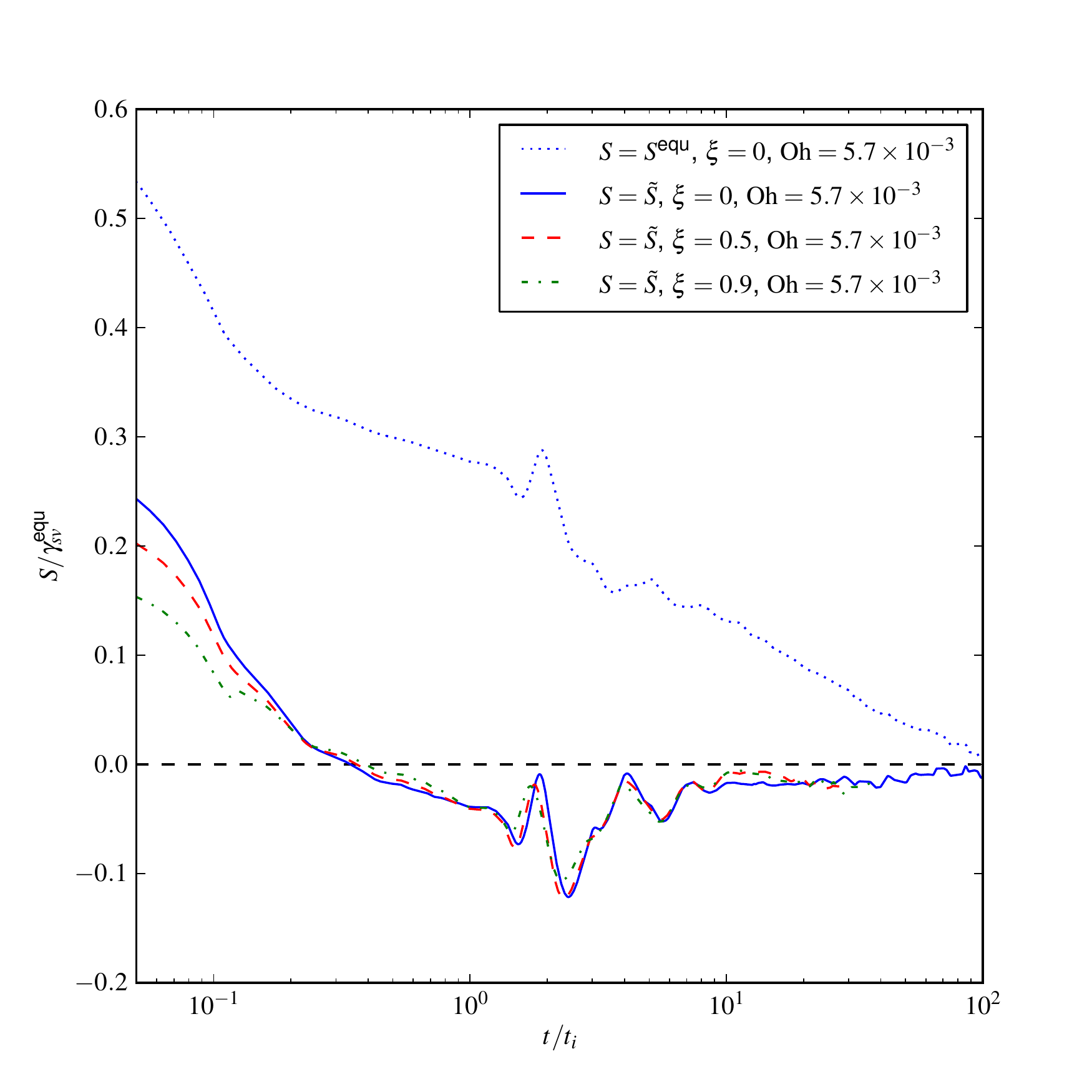}}\hfill}
  \caption{The scaled spreading coefficient versus scaled time for various values
    of $\xi$.}
  \label{fig:timeVAngle-X1}
\end{figure}

Using our definition of $\tilde\gamma$ and the apparent contact angle, $\theta$,
we can now calculate dynamic values of both $\tilde{S}$ and $S^{\text{equ}}$,
which are presented in figure~\ref{fig:timeVAngle-X1}. The curves decrease
rapidly from their maximum value and become negative at about $t=t_i$ and then
oscillate in conjunction with the triple-line radial position
oscillations. Eventually, the values of $\tilde{S}$ become quite small ($<10\%$
of its original value for $\xi=0$) although the drop is still spreading.
Assuming $\tilde{S}$ quantifies the driving force for spreading, then the
differences in $\tilde{S}$ that occur for different values of $\xi$ at early
times may explain both the deviations observed in the spreading extent during the
inertial regime ($t< t_i$) and the deviations in the oscillation amplitudes in
figure~\ref{fig:dropRadius}. The small values of $\tilde{S}$ when compared with
$S^{\text{equ}}$ at late times suggest that the spreading has become quasi-static
in nature and is bound to the evolving values of the $\tilde{\gamma}$. The
evolution of the $\tilde{\gamma}_{sl}$ occurs on a time scale associated with
$t_{\text{diff}}$ while the hydrodynamic adjustment of the contact angle occurs
on a time scale associated with $t_i$. Thus, the contact angle can adjust rapidly
to balance the horizontal forces and suggests that the spreading is limited by
interface equilibration at late times.

\subsection{Dissipation analysis\label{section:dissipation}}

Much of the literature surrounding droplet spreading is concerned with
characterizing dissipation mechanisms from the point of view of an irreversible
thermodynamic process~\cite{ISI:000225453200025, RevModPhys.57.827,
  BrochardWyart19921}.  In this spirit, this section provides an analysis of the
entropy production, yielding the magnitudes of the various dissipation mechanisms
in our model, which should, in turn, provide guidance on the formulation of
simplified models.  The expression used here for the total entropy production
rate is given by~\cite{sekerka2002phase},
\begin{equation}
  \dot{S}_{\text{PROD}} = \frac{M}{T^2} | \partial_j \left( \mu_1^{\text{NC}} - \mu_2^{\text{NC}} \right) |^2
  + \frac{M_{\phi}}{T^2} \left( \frac{\partial f}{\partial \phi} - \epsilon_{\phi} T \partial_j^2 \phi \right)^2
  + \frac{\nu}{2 T} \left( \partial_i u_k + \partial_k u_i \right) \partial_i u_k  
\label{eqn:entropy}
\end{equation}
where each term in the sum is a distinct dissipation mechanism
(diffusion, solid interface relaxation, and viscous flow).

The comprehensive overview of wetting by de Gennes~\cite{RevModPhys.57.827}
identified three main mechanisms for dissipation in spreading droplets: a viscous
dissipation concerned with the ``rolling motion'' of the fluid within
\unit{100}{\mu\meter} of the triple line, a viscous dissipation in the precursor
film and a highly localized dissipation at the triple line associated with
``triple-line friction''. In the present work, the precursor film is absent,
however, both viscous dissipation in the bulk fluid and local triple-line
dissipation are present, but are conflated within the viscous dissipation term in
Eq.~\eqref{eqn:entropy}. In most models of droplets spreading, the chosen model
for slip relaxation at the triple line influences the underlying dissipation
mechanism for the spreading droplet. For example, a molecular kinetics model of
slip generally implies a local triple-line dissipation, while a hydrodynamic
model of slip, such as Cox's model~\cite{ISI:A1986D634200007} or Tanner's
law~\cite{ISI:A1979HN64000009}, both examples of de Gennes' ``rolling motion'',
implies non-localized dissipation~\cite{ISI:000225453200025, BrochardWyart19921}.
We are reminded that this model employs diffuse interfaces, and thus no explicit
slip condition is postulated, but such slip is a direct consequence of the model.

\begin{figure}
  {\hfill\scalebox{0.5}{\includegraphics{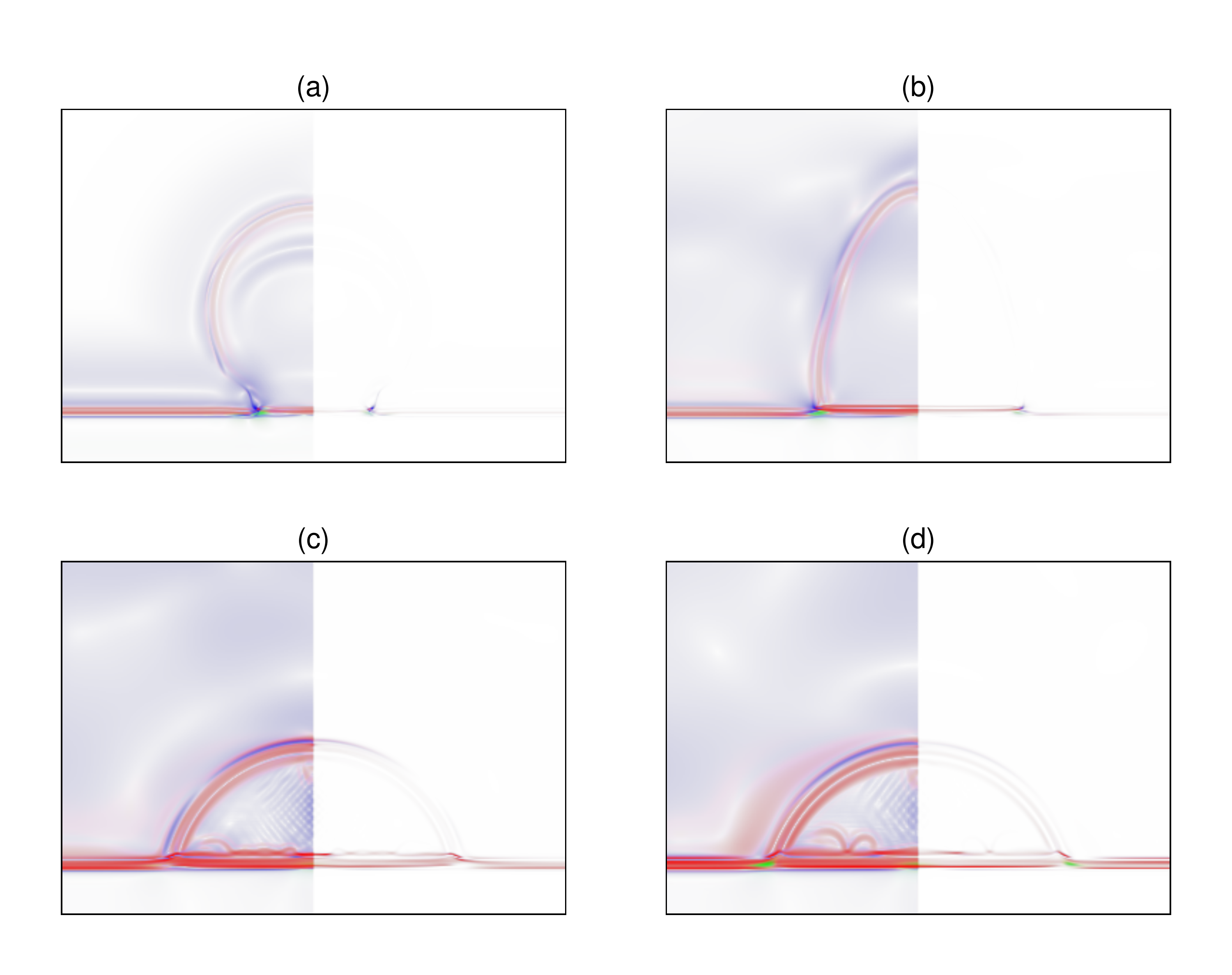}}\hfill}
  \caption{Contour plots of the entropy production rate at (a) $t = 0.1 t_i$, (b)
    $t = t_i$, (c) $t = 10 t_i$ and (d) $t = 20 t_i$. The color intensity
    represents the magnitude of either $\sqrt[8]{\dot{S}_{\text{PROD}}}$ (less
    focused) on the left panel or $\sqrt{\dot{S}_{\text{PROD}}}$ (more focused)
    on the right panel.  The colors represent the specific entropy production
    mechanism given by the terms in Eq.,~\eqref{eqn:entropy} (diffusive, phase
    field, viscous), with red, green and blue representing the first (diffusion),
    second (solid interface relaxation) and third (viscous flow) terms,
    respectively.}
  \label{fig:entropy}
\end{figure}

Figure~\ref{fig:entropy} presents color contour plots of the entropy production
rates at various times. The plots show the magnitude, location and mechanism of
entropy production for the non-dissolutive case (the dissolutive cases are only
slightly different). The color mapping is rescaled in figure~\ref{fig:entropy}
based on the $\max \left( \dot{S}_{\text{PROD}} \right)$ value for each
image. For example, the total entropy production rate in figure~\ref{fig:entropy}
(d) is only 0.4\% of the value in figure~\ref{fig:entropy} (a).  If we were
considering a non-isothermal system, there would be a further term in
expression~\ref{eqn:entropy} containing temperature gradients, an effect not
considered in this work.

At very early times (not shown), the entropy production is highly localized at
the solid-fluid interface region as $\phi$ locally equilibrates. Subsequently
(not shown), pressure waves are observed as the liquid-vapor interface
equilibrates, and viscosity is the dominant mode of dissipation. By $t=0.1 t_i$,
the pressure waves have mostly subsided and the spreading is well under way. At
this stage, the dominant dissipation mechanism remains viscous but is now highly
localized at the triple-line. As the inertial time scale is approached in
figure~\ref{fig:entropy} (b), the dominant mechanism alternates between diffusive
and viscous as the droplet oscillates during the $t_i < t < 10 t_i$ stage. The
viscous dissipation remains highly localized at the triple line, while the
diffusive dissipation mostly occurs in the solid-liquid interface with some
occurring along the solid-vapor interface. This correlates with
figure~\ref{fig:gamma-205}, which shows that the solid-liquid interface is far
from local equilibrium until much later times. At later times
(figure~\ref{fig:entropy} (c)), dissipation is mainly due to local interface
equilibration along the solid-liquid and solid-vapor interface regions. The
proportion of the numerically integrated value of $\int \dot{S}_{\text{PROD}} dV$
for each term in Eq.~\eqref{eqn:entropy} (diffusive, phase field, viscous) is (a)
(0.51, 0.06, 0.43), (b) (0.73, 0.04, 0.23), (c) (0.84, 0.02, 0.14) and (d) (0.85,
0.07, 0.08) for each subplot in figure~\ref{fig:entropy}. These proportions
demonstrate the growing influence of diffusive dissipation and the reduction in
viscous dissipation as the system transitions from the inertial regime to the
diffusive regime.

\subsection{Remarks}

The temporal adjustment to the equilibrium interface profiles is extremely
complex and intimately related to the interface width and the interpolated values
of the dynamic coefficients ($\nu$ and $\bar{M}$), see Eqs.~\eqref{eqn:mbar}
and~\eqref{eqn:nu}. The choice for the interpolation parameter $a$ in
Eqs.~\eqref{eqn:mbar} and~\eqref{eqn:nu} biases the coefficients to have values
close to the bulk fluid values in the interface region facilitating the fastest
interface dynamics possible within the bounds set by the bulk values. The
parameter $a$ is tuned to a value of 4, as larger values do not increase the
interface equilibration rate while smaller values considerably reduce the
equilibration rate.

The equilibration of the density and phase field interface profiles is fast
compared to that of the concentration field. The interface profile of the density
field, $\rho$, is adjusted rapidly by hydrodynamics alone, while the interface
profile of the concentration field, $\rho_1 / \rho$, requires inter-diffusion
between the bulk phases and the interface regions. This compositional relaxation
could, in principle, be as slow as the diffusion time scale, $t_{\text{diff}}$
(see Table~\ref{tab:timescales}), although the connection is imprecise, as this
quantity is associated primarily with the motion of the interface due to
dissolution (melting) rather than the relaxation of compositional profiles within
the interface.  The solid interfaces equilibrate slowly, compared to the
liquid-vapor interface, as seen in figure~\ref{fig:gamma-205}. We expect that the
observed interface relaxation time is unrealistic, when compared with
experimental studies of metallic systems, as our chosen interface width of
$\delta=\unit{100}{\nano\meter}$ is much larger than the
$\delta\approx\unit{1}{\nano\meter}$ typical of metals.  This is a shortcoming of
this treatment, and results in an unphysical time scale for local interface
equilibration. Further analysis of the relationship between $\delta$ and the
equilibration rate is required, though this analysis is beyond the scope of this
work. The limitation of requiring $\delta / R \approx 0.1$ imposed by the
available computation resources does not detract from the analysis presented in
this section with respect to the reduced spreading when $\xi$ is increased, the
qualitative description of the spreading regimes and oscillations, or the
quantitative comparisons with experiments.

\section{Conclusion \label{section:conclusion}}

This paper presents results from a model of dissolutive spreading simulated in a
parameter regime where inertial effects are initially dominant. The triple-line
motion demonstrates good agreement with the $O(t^{-\nicefrac{1}{2}})$ inertial
spreading rate at early times. The model also generates oscillations
characteristic of the transition from inertial to viscous or diffusive
spreading. Subsequent analysis indicates that a force balance involving the
instantaneous interface energies evaluated using the expression in
Eq.~\eqref{eqn:surfaceTensionIntegral} can explain the variation in spreading
between the hydrodynamic and dissolutive cases. At late times, after inertial
effects have ceased, the contact angle derived from the instantaneous interface
energies is within \unit{0.005}{\radian} of the measured contact angle suggesting
that the local interface equilibration mechanism is controlling the
spreading. Analysis of the dissipation mechanism via the entropy production
expression demonstrates that dissipation occurs at the triple line during the
inertial stage, but transitions to the solid-fluid interfaces during the
oscillatory stage consistent with the instantaneous interface energy analysis.
Overall, the simulation results show good quantitative and qualitative agreement
with a number of experimental results when time is scaled with the inertial time
scale.

Modeling droplets that have both a realistic interface width and include inertial
effects is impractical with current computational resources (at least for the
model presented herein) and may require years of real time computation on large
parallel clusters.  In this work, to reduce the required compute time, the use of
a realistic interface width has been sacrificed in order to preserve the inertial
effects. This has the consequence of increasing the simulation time required for
the local equilibration process across the solid-fluid interface as discussed in
section~\ref{subsection:interface}. Although, this process has a longer duration
than physically appropriate in the present work, a time regime over which the
controlling mechanism for spreading is the local interface equilibration may be
entirely physical. It is noted in Protsenko~\textit{et
  al.}~\cite{ISI:000258573900040} that the diffusive stage may occur in two
separate parts. The first part is surmised to be the solid-liquid interface
equilibration process and takes approximately an order of magnitude longer than
the inertial time scale, which is faster than occurs here, but very similar in
nature. The second part is the melting of the substrate, which is included in
this model, but not observed as it occurs over a time scale longer than the total
duration of a typical simulation.

Further work may involve both direct comparison with molecular kinetics theory
and more detailed analysis of the impact of the interface width on the spreading
dynamics.

\section{Acknowledgements}

The authors would like to acknowledge the contributions of Dr. Jonathan E. Guyer
and Dr. Walter Villanueva for their help and guidance in implementing the
numerical model and analyzing the numerical data, and Dr. Edmund B. Webb for
insightful commentary and help in setting this work in the proper context.

\bibliography{paper}

\appendix
\section{Derivation of the Governing Equations
  \label{appendix:derivation}}

In this section the underlying thermodynamic and constitutive relationships
required for the derivation of Eqs.~\eqref{eqn:diffusion},~\eqref{eqn:phase}
and~\eqref{eqn:momentum2} are presented.

As previously outlined, the fluid phases are represented by a binary, van der
Waals equation of state and the solid phase is represented by a simple linear
compressive and tensile equation of state that ignores all shear stress. The van
der Waals equation of state is given by,
\begin{equation}
  \left( P - \frac{n^2}{V^2} \left(e_1 X_1 + e_2 X_2 \right) \right) \left( V - \bar{v} n \right) = n R T
\label{eqn:vanderwaals}
\end{equation}
where $X_1$ and $X_2$ are the concentrations of each component, $n$ is the number
of moles and $V / n = m / \rho$. All other parameters used in
Eq.~\eqref{eqn:vanderwaals} are defined in section~\ref{section:equations}.
Eq.~\eqref{eqn:vanderwaals} can be related to the ideal gas law, but has modified
pressure and volume terms to account for the long range attraction of molecules
and volume exclusion, respectively~\cite{kittel1980thermal,
  plischke1994equilibrium}. The solid equation of state is given by,
\begin{equation}
  P V_s = 2 B n \frac{V_s - V}{V_s}
\label{eqn:solidstate}
\end{equation}
where $V_s / n = m / \rho_s^{\text{ref}}$. The Helmholtz free energies given in
Eqs.~\eqref{eqn:fluidfreeenergy} and~\eqref{eqn:solidFreeEnergy} are derived
from~\eqref{eqn:vanderwaals} and~\eqref{eqn:solidstate}, respectively, using the
thermodynamic identities given in Eqs.~\eqref{eqn:mu1},~\eqref{eqn:mu2}
and~\eqref{eqn:P}. In order to derive
Eqs.~\eqref{eqn:diffusion},~\eqref{eqn:phase} and~\eqref{eqn:momentum2}, it is
necessary to postulate a form for the free energy functional.

As in reference~\cite{Bi199895}, standard non-classical
diffuse interface expressions for $\rho_1$, $\rho_2$ and $\phi$ are used, which
results in a functional of the form,
\begin{equation}
  F = \int \left[ f + \frac{\epsilon_{\phi} T}{2} |\nabla \phi|^2 + \frac{\epsilon_{1} T }{2} |\nabla \rho_1|^2 +
    \frac{\epsilon_{2} T}{2} |\nabla \rho_2|^2
  \right] dV
  \label{eqn:functional}
\end{equation}
Using standard dissipation arguments~\cite{Bi199895}, Eqs.~\eqref{eqn:diffusion}
and~\eqref{eqn:phase} are derived using,
\begin{equation*}
  \frac{\partial \phi}{\partial t} + u_j \partial_j \phi  = - M_{\phi} \frac{\delta F}{\delta \phi}
\end{equation*}
and
\begin{equation*}
  \frac{\partial \rho_1}{\partial t} + \partial_j \left( u_j \rho_1 \right) = -\partial_j J_{1j}
\end{equation*}
and similarly for component 2. The fluxes are given by,
\begin{equation*}
  J_{1j} = -J_{2j} = - M \partial_j \left( \frac{ \mu_1^{NC} - \mu_2^{NC}}{T} \right)
 \end{equation*}
where,
\begin{equation*}
  \mu_1^{NC} = \frac{\delta F}{\delta \rho_1}  
\end{equation*}
and
\begin{equation*}
  \mu_2^{NC} = \frac{\delta F}{\delta \rho_2}  
\end{equation*}
The form of the stress tensor required to derive the momentum equation is given
by,
\begin{equation*}
  \sigma_{ij} = \nu \left( \partial_j u_i + \partial_i u_j \right) + t_{ij}
\end{equation*}
using the standard assumption that the bulk viscosity, $\lambda$, is related to
the shear viscosity via $\lambda = -\frac{2}{3} \nu$. The tensor, $t_{ij}$, is
derived from a conservation law ($\partial_j t_{ij} = 0$) based on Noether's
theorem~\cite{ISI:000071880700007}. The expression for $t_{ij}$ is given by,
\begin{equation}
  t_{ij} =  g^{NC} \delta_{ij}  - \partial_j \rho \frac{\partial g^{NC}}{ \partial \left( \partial_i \rho \right)} 
  \label{eqn:tij}
\end{equation}
where
\begin{equation}
  g^{NC} = f^{NC} + \rho_1 \lambda_1 + \rho_2 \lambda_2
  \label{eqn:gnc}
\end{equation}
The non-classical Gibbs free energy, $g^{NC}$, is the form of the free energy
that includes Lagrange multipliers for conservation of species 1 and 2.  The
Lagrange multipliers for each species are equal to $\lambda_1 = -\mu_1^{NC}$ and
$\lambda_2 = -\mu_2^{NC}$ in equilibrium using the variational derivative of $
\int g^{NC} dV$ with respect to $\rho_1$ and $\rho_2$. Using Eqs.~\eqref{eqn:tij}
and~\eqref{eqn:gnc} the form for $\partial_i t_{ij}$ used in
Eq.~\eqref{eqn:momentum2} can be derived,
\begin{equation}
  \partial_j t_{ij} = -\rho_1 \partial_i \mu_1^{NC} - \rho_2 \partial_i \mu_2^{NC} - \partial_i \phi \frac{\delta F}{\delta \phi}
\end{equation}

\section{Numerical Approach \label{appendix:numericalApproach}}

In general, even for compressible systems, many conventional algorithms use the
pressure field as the independent variable rather than the density field. This
approach is thought to have more robust convergence
properties~\cite{ferziger:1996} at low Mach numbers due to the weak dependence of
pressure gradients on density, but the convergence properties deteriorate at
higher Mach numbers. In this work, due to the non-trivial nature of the
pressure-density relationship, an inversion of this relationship would be
impractical and it is more natural to solve for the density field rather than the
pressure field. Due to the mesh collocation of the density and velocity fields,
an interpolation scheme, known as Rhie-Chow interpolation~\cite{rhie:1983}, is
employed to ensure adequate velocity-pressure coupling.

The calculation of triple-line velocities is necessarily noisy, with fluctuations
on a timescale of $\Delta x/U$, where $\Delta x$ is the fixed grid spacing.  In
figure~\ref{fig:triplePointVelocityVtime}, the curves are constructed using a 20
point boxcar (equally-weighted) averaging scheme collected at every 10 time steps
during the simulation.  We note that the sign changes in the blue curve
($t_i<t<10t_i$) in figure~\ref{fig:triplePointVelocityVtime} correspond to the
triple-line oscillations, and are {\sl not} due to the averaging scheme. The
velocity fluctuations will be small when $U$ is large.  Indeed, at early times,
when $t / t_i < 1$, $U$ is relatively large and the results are smooth. At later
times, when $t / t_i > 10$, the averaging scheme does not smooth out the noise,
as the spreading rate is greatly reduced. This can be seen in the noisy behavior
at long times for the $\operatorname{Oh}=5.7\times10^{-3}$ curve (blue) in
figure~\ref{fig:triplePointVelocityVtime}.  The noise in the low velocity regime
of figure~\ref{fig:CaVAngle} also reflects this behavior.

The measurements for $\theta$ are calculated using the tangent to the
liquid-vapor interface at a distance of 1.3 $\delta$ from the triple-line
location. In general, this distance results in a reasonable approximation to the
apparent contact angle.

\subsection{Parasitic Currents}

Parasitic currents are a common source of numerical errors when computing flows
with interface energy driving forces that have large $\operatorname{Ca}$.
Typically, for the systems of interest in this paper, $\operatorname{Ca} \approx
10^{-2}$, but parasitic velocities were still found to be a source of numerical
error, particularly when trying to evaluate equilibrium solutions. Parasitic
currents are characterized by quasi-steady flow fields that do not dissipate over
time despite the system reaching equilibrium in all other respects. This can
result in equilibrium errors in both the density and concentration
fields. Jamet~\textit{et al.}~\cite{jamet:2002} as well as other researchers have
demonstrated that parasitic currents can be eliminated by recasting the momentum
equation in a form that only conserves momentum to the truncation error of the
discretization rather than machine precision. The form of the momentum equation
that eliminates parasitic currents is written in terms of the chemical potentials
and is given by,
\begin{equation}
  \frac{\partial \left( \rho u_i \right) }{\partial t} +
  \partial_j \left( \rho u_i u_j \right) =
  \partial_j \left( \nu \left[ \partial_j u_i + \partial_i u_j \right]
  \right) -
  \rho_1 \partial_i \mu_1^{NC} - \rho_2 \partial_i \mu_2^{NC}
  \label{eqn:momentum-energy-conserving}
\end{equation}
for binary liquid-vapor system. The discretized form of
Eq.~\eqref{eqn:momentum-energy-conserving} is known as an energy conserving
discretization in contrast to the momentum conserving discretization, which
results when the momentum equation is written in terms of the pressure (see
Eq.~\eqref{eqn:momentum-dimensionless}).

\subsection{Convergence}

Some simulations in this paper are tested for convergence with grid sizes of
180$\times$125, 360$\times$250 and 720$\times$500 using the triple-line and drop
height positions against time as the metrics for convergence. Production runs for
the results presented use 360$\times$250 grids. Details of these convergence
tests can be found in~\cite{reactiveWettingInstallation}. Convergence at the
$n^{\text{th}}$ time-step is achieved when the $k^{\text{th}}$ iteration within
the time step satisfies the residual condition $ \beta_n^k / \beta_n^0 < 1 \times
10^{-1}$ for each of the equations where $\beta_n^k$ is the $L_2$-norm of the
residual at the the $k^{\text{th}}$ iteration of the $n^{\text{th}}$ time
step. Further decreases in the residual make little difference to the dynamic
positions of the drop height and triple-line. Numerical calculations indicate
that, in the course of a simulation, $\operatorname{Ma}$ ranges from values that
require compressible flow solvers (density based with $\operatorname{Ma} >
2\times 10^{-1}$) to values for which compressible flow solvers have trouble with
accuracy and convergence for traditional segregated solvers ($\operatorname{Ma} <
2\times 10^{-1}$). The shift to low $\operatorname{Ma}$ generally occurs when the
system is quite close to equilibrium and is not believed to affect the dynamic
aspects of the simulation, which are of most interest in this paper. In general,
for low Mach number flows, preconditioners are used to improve the convergence
properties of segregated solvers. In this work, it was found that using a coupled
solver along with a suitable preconditioner greatly improved the convergence
properties.  The preconditioners are available as part of the Trilinos software
suite~\cite{Trilinos-Overview}. The coupled convergence properties can be further
improved by employing physics based preconditioners that change the nature of the
equations based on the value of $\operatorname{Ma}$~\cite{keshtiban:2004}, but
are not used in this work.

\begin{table}
  \begin{tabular}{c|r@{.}ll}
    Parameter        & \multicolumn{2}{c}{Value}                    & Unit                                                              \\
    \hline
    $\nu_f$         & 2 &0$\times$10$^{-3}$                         & \kilogram\per(\second\cdot\metre)                                 \\
    $\nu_s$         & 2 &0$\times$10$^{4}$                          & \kilogram\per(\second\cdot\metre)                                 \\
    $\epsilon_1$     & 2 &0$\times$10$^{-16}$                        & \power{\metre}{7}\per(\kelvin\cdot\kilogram\cdot\square\second)     \\
    $\epsilon_2$     & 2 &0$\times$10$^{-16}$                        & \power{\metre}{7}\per(\kelvin\cdot\kilogram \cdot\square\second)     \\
    $T$              & 6 &5$\times$10$^{2}$                          & \kelvin                                                           \\
    $m$              & 1 &18$\times$10$^{-1}$                        & \kilogram\per\mole                                                 \\
    $R$              & 8 &31                                        & \joule\per(\kelvin\cdot\mole)                                      \\
    $v_a$            & 1 &0                                          &                                                                   \\
    $e_1$            & -4&56$\times$10$^{-1}$                        & \joule\cdot\cubic\metre\per\square\mole                           \\
    $e_2$            & -4&56$\times$10$^{-1}$                        & \joule\cdot\cubic\metre\per\square\mole                           \\
    $\bar{v}$        & 1 &3$\times$10$^{-5}$                         & \cubic\metre\per\mole                                              \\
    $A_1$            & 2 &83$\times$10$^{4}$                         & \joule\per\mole                                                    \\
    $A_2$            & 5 &64$\times$10$^{4}$                         & \joule\per\mole                                                    \\
    $\rho_s^{\text{ref}}$ & 7 &84$\times$10$^{-5}$                      & \kilogram\per\cubic\metre                                         \\
    $B$              & 2 &02$\times$10$^{5}$                         & \joule\per\mole                                                    \\
    $W$              & 1 &27$\times$10$^{5}$                         & \newton\per\square\metre                                           \\
    $\epsilon_{\phi}$ & 1 &0$\times$10$^{-9}$                         & \newton\per\kelvin                                                 \\
    $M_{\phi}$        & 1 &0$\times$10$^{4}$                          & \kelvin\cdot\square\metre\per(\newton\cdot\second)                 \\
    $\bar{M}_f$      & 1 &0$\times$10$^{-7}$                         & \kilogram\cdot\second\cdot\kelvin\per\cubic\metre                   \\
    $\bar{M}_s$      & 1 &0$\times$10$^{-11}$                          &   \kilogram\cdot\second\cdot\kelvin\per\cubic\metre \\
    $R_0$               & 1 &0$\times$10$^{-6}$                          &   \metre \\
    $\delta$            & 1 &0$\times$10$^{-7}$                          &   \metre \\
    $\rho_l^{\text{equ}}$ & 7 &35$\times$10$^{3}$                          &   \kilogram\per\cubic\metre
  \end{tabular}
  \caption{Various parameter values.}
  \label{tab:2Dsimulation}
\end{table}

\begin{table}
  \begin{tabular}{lll|r@{.}l}
    Time scale                  & Symbol             & Expression                                      & \multicolumn{2}{c}{Value (s)}        \\
    \hline
    capillary                   & $t_c$              & $\nu_f R_0/ \gamma_{lv}$                         & 1&05$\times10^{-10}$                   \\
    phase field                 & $t_{\phi}$          & $\delta^2 / \epsilon_{\phi} M_{\phi}$           & 1&0$\times10^{-9}$                      \\
    inertial                    & $t_i$              & $\sqrt{\rho_l^{\text{equ}} R_0 / \gamma_{lv}}$   & 1&97$\times10^{-8}$                     \\ 
    convection                  & $t_a$              & $R_0 / U$                                        & 1&97$\times10^{-8}$                    \\
    viscous                     & $t_{\nu}$              & $\rho_l^{\text{equ}} R_0^2 / \nu_l$          & 3&68$\times10^{-6}$                    \\
    interface diffusion         & $t_{\text{diff}}$    & $\delta^2 / 4 K^2 D_f$                         & 7&69$\times10^{-4}$                     \\
    bulk diffusion              & $t_{\text{d}}$      & $R_0^2 / D_f$                                   & 1&04$\times10^{-3}$                     \\
    solid deformation           & $t_s$              & $\delta \nu_s / \gamma_{lv}$                     & 1&05$\times10^{-2}$                    \\
    instantaneous convection    & $t_a^*$            & $R_0 / U^*$                                      & &                                   \\
  \end{tabular}
  \caption{Complete list of time scales referred to in this paper.}
  \label{tab:timescales}
\end{table}

\begin{table}
  \begin{tabular}{lll|r@{.}l}
    Parameter                           &        Symbol                          & Expression                                             &\multicolumn{2}{c}{Value}      \\
    \hline
    Peclet number                       & $\operatorname{Pe}$                    & $U R_0 / D_f = t_d / t_a$                                       &5&$31\times10^{4}$               \\ 
    Reynolds number                     & $\operatorname{Re} $                   & $U R_0 \rho_l^{\text{equ}} / \nu_f = t_{\nu} / t_a $            &1&$87\times10^{2}$               \\
    Weber number                        & $\operatorname{We}$                    & $\operatorname{Re} \, \operatorname{Ca} = t_{\nu} t_c / t_a^2$  &1&0                             \\
    Mach number                         & $\operatorname{Ma}$                    &  $U / c$                                                        &5&$72\times10^{-2}$               \\  
    Unnamed                             & $\operatorname{Q}$                     & $ m \gamma_{lv} / R T \rho_l^{\text{equ}} R_0 $                 &5&$64\times10^{-2}$                \\
    effective dimensionless slip length & $\lambda$                              & $\delta / R_0$                                                  &5&$0\times10^{-2}$          \\ 
    Capillary number                    & $\operatorname{Ca}$                    & $U \nu_f / \gamma_{lv} = t_c / t_a$                             &5&$35\times10^{-3}$             \\ 
    Ohnesorge number                    & $\operatorname{Oh}$                    & $\sqrt{\operatorname{Ca} / \operatorname{Re}} = t_c / t_i$      &5&$35\times10^{-3}$              \\
    instantaneous Reynolds number       & $\operatorname{Re}^*$                  &  $U^* R_0 \rho_l^{\text{equ}} / \nu_f = t_{\nu} / t_a^*$        &&                                \\
    instantaneous Capillary number      & $\operatorname{Ca}^*$                  & $U^* \nu_f / \gamma_{lv} = t_c / t_a^*$                         &&                            \\ 
  \end{tabular}
  \caption{Relevant dimensionless numbers.}
  \label{tab:dimensionlessnumbers}
\end{table}

\end{document}